\newif\ifsubmode
\submodefalse

\newif\ifprintfig
\printfigtrue

\newif\ifemulate
\emulatetrue

\ifsubmode
  \documentclass[12pt,preprint]{aastex}
  \received{}
  \accepted{}
  \journalid{}{}
  \articleid{}{}
\else
   \documentclass[twocolappendix]{emulateapj}
   \submitted{{\it To be submitted for publication in ApJ}}
\fi

\newcommand{\kms}{\,km~s$^{-1}$}

\def\lesssim{\mathrel{\hbox{\rlap{\hbox{\lower4pt\hbox{$\sim$}}}\hbox{$<$}}}}
\def\gtrsim{\mathrel{\hbox{\rlap{\hbox{\lower4pt\hbox{$\sim$}}}\hbox{$>$}}}}
\def\spose#1{\hbox to 0pt{#1\hss}}
\def\simlt{\mathrel{\spose{\lower 3pt\hbox{$\mathchar"218$}}
     \raise 2.0pt\hbox{$\mathchar"13C$}}}
\def\simgt{\mathrel{\spose{\lower 3pt\hbox{$\mathchar"218$}}
     \raise 2.0pt\hbox{$\mathchar"13E$}}}

\slugcomment{Accepted for publication in the Astrophysical Journal}

\shorttitle{Cosmic Ray Sampling of ISM}
\shortauthors{Boettcher, Zweibel, Yoast-Hull, and Gallagher}

\begin{document}

\title{Cosmic Ray Sampling of a Clumpy Interstellar Medium}

\author{Erin Boettcher$^{1}$, Ellen G. Zweibel$^{1,2,3}$, Tova
  M. Yoast-Hull$^{2,3}$, and J. S. Gallagher III$^{1}$}

\affiliation{$^{1}$Department of Astronomy, University of Wisconsin, Madison,
  WI 53706, USA; \texttt{boettche@astro.wisc.edu}}
\affiliation{$^{2}$Department of Physics, University of Wisconsin, Madison, WI 53706, USA}
\affiliation{$^{3}$Center for Magnetic Self-Organization in Laboratory and Astrophysical Plasmas, University of Wisconsin, Madison, WI 53706, USA}


\ifsubmode\else
  \ifemulate\else
     \clearpage
  \fi
\fi


\ifsubmode\else
  \ifemulate\else
     \baselineskip=14pt
  \fi
\fi

\begin{abstract}

  How cosmic rays sample the multi-phase interstellar medium (ISM) in
  starburst galaxies has important implications for many science goals,
  including evaluating the cosmic ray calorimeter model for these systems,
  predicting their neutrino fluxes, and modeling their winds. Here, we use
  Monte Carlo simulations to study cosmic ray sampling of a simple, two-phase
  ISM under conditions similar to those of the prototypical starburst galaxy
  M82. The assumption that cosmic rays sample the mean density of the ISM in
  the starburst region is assessed over a multi-dimensional parameter space
  where we vary the number of molecular clouds, the galactic wind speed, the
  extent to which the magnetic field is tangled, and the cosmic ray injection
  mechanism. We evaluate the ratio of the emissivity from pion production in
  molecular clouds to the emissivity that would be observed if the cosmic rays
  sampled the mean density, and seek areas of parameter space where this ratio
  differs significantly from unity. The assumption that cosmic rays sample the
  mean density holds over much of parameter space; however, this assumption
  begins to break down for high cloud density, injection close to the clouds,
  and a very tangled magnetic field. We conclude by evaluating the extent to
  which our simulated starburst region behaves as a proton calorimeter and
  constructing the time-dependent spectrum of a burst of cosmic rays.
  
\end{abstract}


\section{Introduction}\label{sec_intro}

Starburst galaxies are complex environments with intense star formation, 
``clumpy,'' multi-phase interstellar gas, supernovae-driven winds, and
tangled magnetic fields. It is remarkable, then, that despite drastically
different environments, both quiescent and star-forming disk galaxies have a
well-established linear correlation between their far-infrared (FIR) and radio
luminosities \citep{Helou1985}. In these systems, the FIR-radio luminosity
correlation suggests a fundamental relationship between the star formation
processes, resulting in FIR emission from dust heated by young, massive stars,
and the cosmic ray population, producing radio synchrotron emission from
relativistic electrons spiraling along magnetic field lines.

To explain the FIR-radio luminosity correlation, the cosmic ray calorimeter
model for starburst galaxies has been explored \citep[e.g.,][]{Voelk1989,
  Thompson2006, Persic2008, deCeadelPozo2009a, Lacki2011, Paglione2012,
  YoastHull2013}. This model suggests that the energy imparted to cosmic rays
by supernovae is entirely expended through observable emission within the
starburst region. If this model holds for cosmic ray electrons, then the
observed radio synchrotron emission should be solely dependent on the
supernova rate and therefore on the star formation rate (SFR; there is, of
course, also a dependence on magnetic field strength, but this too appears to
be tied to the SFR; \citet{Schleicher2013}). Likewise, if the model holds for
cosmic ray protons, then we expect a similar relationship between the observed
$\gamma$-ray emission from pion production and the SFR.

In order to test the calorimeter model for starburst galaxies, it is necessary
to understand how cosmic rays sample the multi-phase interstellar medium (ISM)
as they undergo radiative, collisional, and advective losses within the
starburst region. In an ISM consisting of cold, molecular clouds embedded in a
hot, low-density medium, cosmic rays will only sample the mean density of the
ISM if they are able to effectively enter the molecular clouds. The leading
theory of cosmic ray acceleration suggests that energetic particles undergo
first-order Fermi acceleration \citep{Fermi1949}, or diffusive shock
acceleration \citep[e.g.,][]{Bell1978, Blandford1978}, at supernova shock
fronts. Additionally, cosmic rays may undergo second-order Fermi acceleration
\citep{Fermi1949} in the diffuse ISM. Both first- and second-order Fermi
acceleration can only occur efficiently in low-density environments, where the
acceleration mechanism can impart energies in excess of ionization losses and
the small-scale magnetic field fluctuations which scatter the particles can
propagate. For these reasons, the majority of cosmic rays are believed to be
injected in the hot, low-density medium.

The hot medium where cosmic rays are injected has a very large filling factor
compared to the molecular gas. Additionally, the hot medium may be actively
advected from the region by a galactic wind. In order for the cosmic rays to
sample the mean density of the ISM, the magnetic field lines along which they
propagate must intersect a sufficiently large number of molecular clouds, and
the cosmic rays must remain within the region for a sufficiently long time
before they are advected away. Therefore, cosmic ray sampling of the mean
density of the ISM is far from a foregone conclusion.

\begin{deluxetable}{lcccc}
\tabletypesize{\scriptsize}
\tablecaption{Properties of M82}
\tablewidth{0pt}
\tablehead{ 
\colhead{Physical Parameters} &
\colhead{Values} &
\colhead{References}
}
\startdata
Distance & 3.9 Mpc & 1\\
Radius SB & 200 pc & 2\\
Scale Height SB & 100 pc & 2\\
Molecular Gas Mass & $\sim 3 \pm 1 \times 10^{8} \; M_{\odot}$ & 3,4\\
Wind Speed (Optical) & $\sim 500 - 600$ \kms & 5\\
Wind Speed (X-Ray) & $\sim 1400 - 2200$ \kms & 6
\enddata
\tablerefs{[1] \citet{Sakai1999}; [2] \citet{ForsterSchreiber2003a}; [3]
  \citet{Naylor2010}; [4] \citet{Wild1992}; [5] \citet{Shopbell1998}; [6] \citet{Strickland2009}.}
\end{deluxetable}

In this paper, we use Monte Carlo simulations to study cosmic ray sampling of
an ISM with properties similar to that of the prototypical starburst galaxy
M82. We selected this galaxy for our study because it has a well-studied
starburst region with well-observed masses of its multi-phase ISM, galactic
wind speed, and supernova rate. In M82, the ISM consists of a hot, diffuse
medium in which dense, warm, ionized gas and dense, cold, molecular clouds are
found \citep[e.g.,][]{Westmoquette2009}. In particular, CO measurements suggest
that the galaxy contains $\sim 3 \pm 1 \times 10^{8} \; M_{\odot}$ of
molecular gas that is largely found within clumpy clouds in the starburst
region \citep[e.g.,][]{Naylor2010}. Additionally, M82 has a well-observed
galactic wind that travels approximately perpendicularly to the galactic plane
and is believed to be primarily driven by supernova shock heating
\citep{Chevalier1985}. Estimates of the outflow velocity range from $\sim 500
- 600$ \kms as indicated by optical emission lines of ionized gas \citep[][and
references therein]{Shopbell1998} to as high as $\sim 1400 - 2200$ \kms as
suggested by X-ray observations \citep{Strickland2009}. Although we use
properties suggestive of M82 in this study (see Table 1 for a summary), our
analysis is applicable to many other such systems.

Here, we evaluate how cosmic ray sampling of the ISM in starburst galaxies is
affected by the number of molecular clouds, the galactic wind speed, the
extent to which the magnetic field is tangled, and the cosmic ray injection
mechanism. In Section 2, we describe our Monte Carlo simulation parameters and
the calculation of an emissivity ratio used to quantify the sampling behavior
of the cosmic ray population. Section 3 details the results of our simulations
and highlights the region of parameter space in which the emissivity ratio is
elevated by a factor of a few. In Section 4, we review our results, compare
them to existing models, and discuss their implications for exploring the
cosmic ray calorimeter model and constructing the time-dependent spectrum of a
burst of cosmic rays.

\section{Monte Carlo Simulations}\label{sec_mcs}

\subsection{Simulation Parameters}\label{sec_parm}

The multi-phase ISM in the starburst region is modeled to consist of a hot,
low-density medium in which cold molecular clouds are embedded. Observations
and numerical simulations suggest that the molecular medium consists of many
clumpy, fragmented clouds \citep[e.g.,][]{Blitz1993, Inoue2012}. Although $\sim
3 \pm 1 \times 10^{8} \; M_{\odot}$ of molecular gas has been detected in the
starburst nucleus \citep[e.g.,][]{Naylor2010}, we take a conservative order of
magnitude estimate of $\sim 1 \times 10^{8} \; M_{\odot}$ of molecular gas
within our spherical starburst region of radius $R = 100$ pc, corresponding to
a mean number density $\langle n\rangle \sim 486$ cm$^{-3}$. Assuming that the
upper limit on the maximum mass of giant molecular clouds in the Milky Way of
$\sim 10^{4} - 10^{6} \; M_{\odot}$ is applicable to M82, this implies on the
order of $N_{c} \sim 10^{3}$ molecular clouds. We hold the total mass of
molecular gas constant while varying the number of clouds from $N_{c} = 200$
to $N_{c} = 3000$ clouds to explore both the nominal case ($N_{c} \sim 3000$)
as well as the limit of very small cloud numbers and very high cloud densities
($N_{c} \sim 200$). Each cloud is taken to have a volume of $V_{c} \sim 27$
pc$^{3}$, and thus the clouds' volume filling factor varies from $\sim 0.1\%$
to $\sim 2\%$.

Note that we are explicitly neglecting the contribution of the hot, low-density
medium to the density sampled by cosmic rays. Due to the diffusive nature of
particle motion along magnetic field lines (which reduces their effective
propagation speed from $c$ to the Alfv\'{e}n speed, $v_{A}$), the contribution
of the low-density gas is greater than the simple ratio $\langle n_{h} \rangle
/ \langle n_{c} \rangle$, where $\langle n_{h} \rangle$ and $\langle n_{c}
\rangle$ are the mean densities of the hot gas and the cold molecular gas,
respectively. We estimate that this is approximately a 10\% effect for the M82
environment. Likewise, warm ($ \sim 10^{4}$ K) ionized gas, in addition to hot
gas and cold molecular gas, is undoubtedly present and would also have a small
effect on the model. We will consider these small contributions in future
work.

\begin{figure}[h]
\epsscale{1.2}\plotone{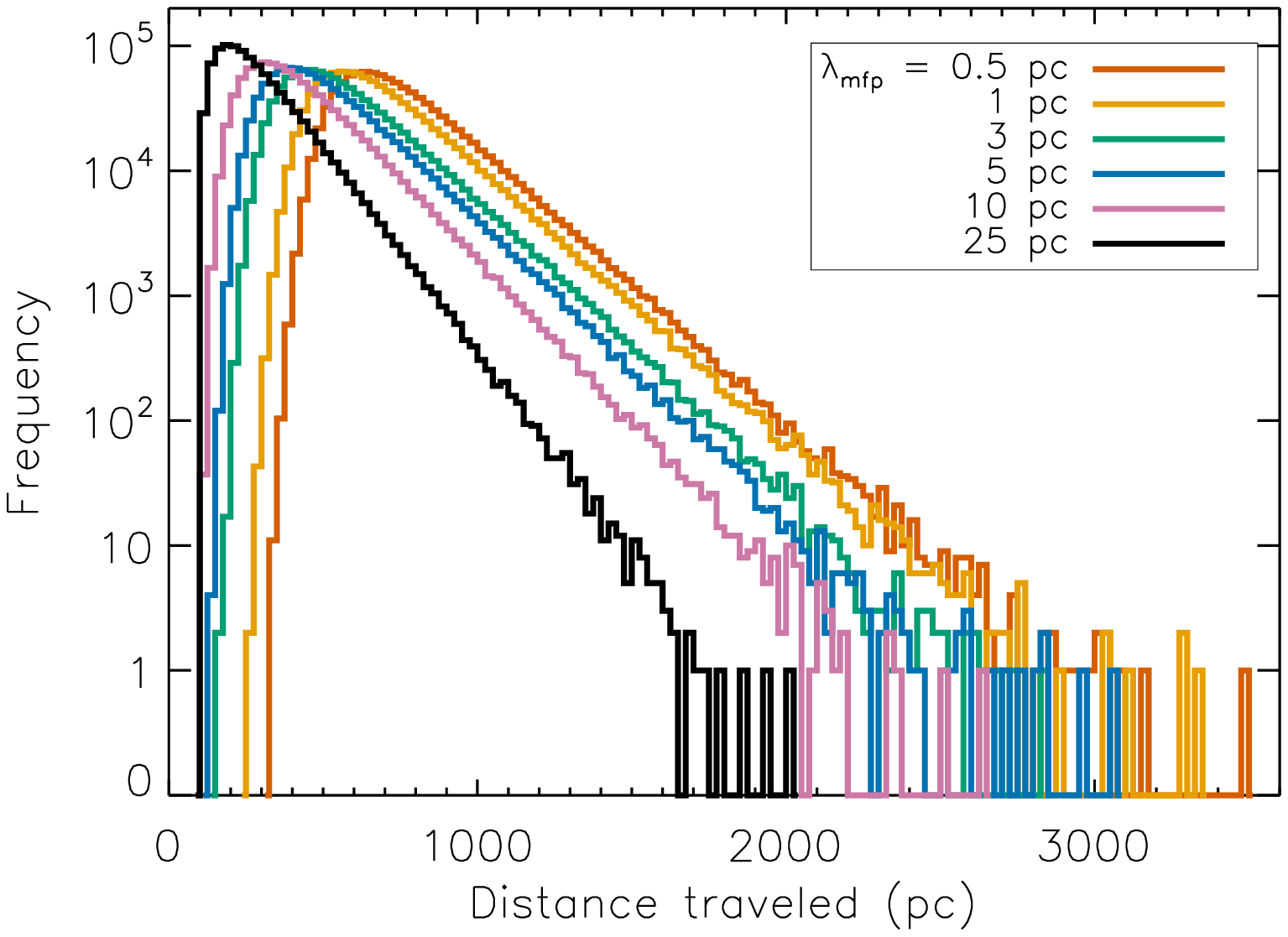}
\epsscale{1.2}\plotone{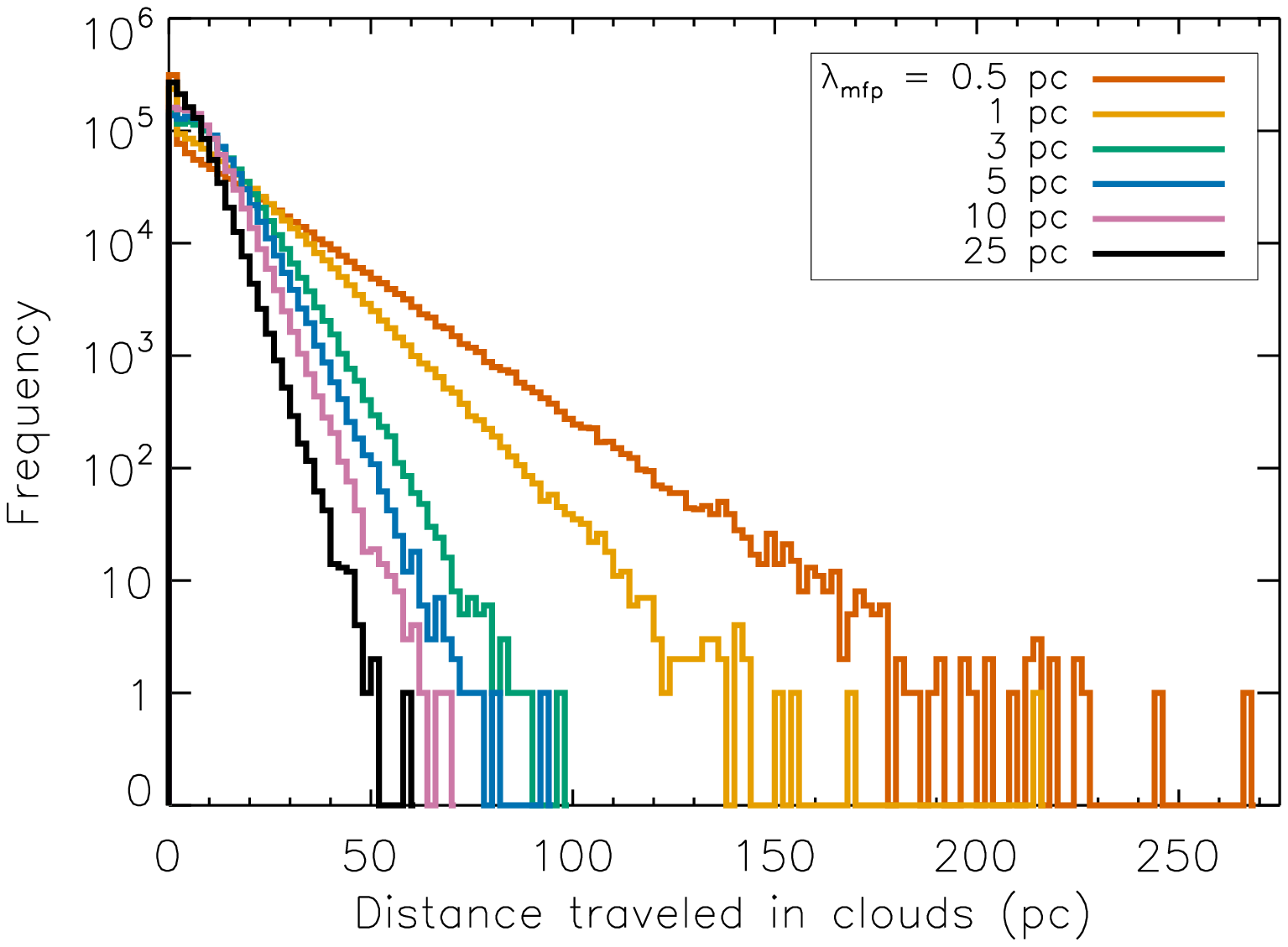}
\epsscale{1.2}\plotone{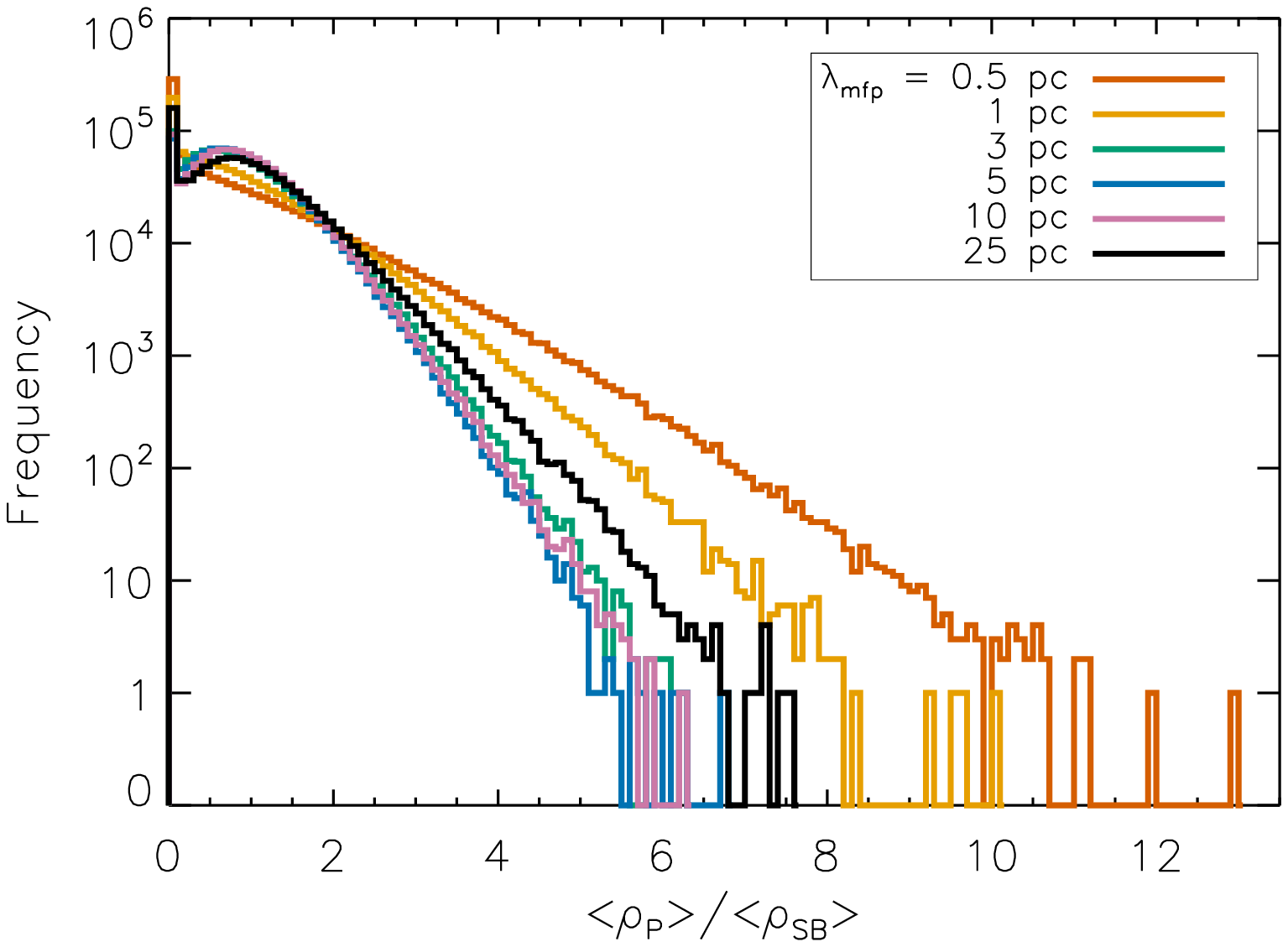}
\caption{Summary of the cosmic ray sampling behavior during a representative
  run of the Monte Carlo code for each $\lambda_{mfp}$ considered. From top to
  bottom, the figures show the distance traveled by particles before leaving
  the starburst region, the distance traveled inside of molecular clouds, and
  the average density sampled normalized to the mean density of the ISM,
  neglecting any contribution from the hot or warm ionized gas. The particles
  show a significant range of sampling behaviors; all density distributions
  show a peak at $\langle \rho_{P} \rangle / \langle \rho_{SB} \rangle = 0$,
  where the particles sample no clouds, as well as varying degrees of a tail
  at $\langle \rho_{P} \rangle / \langle \rho_{SB} \rangle \gtrsim 5$, where
  they sample average densities greater than the mean density of the ISM by as
  much as an order of magnitude. These results were found for injection at the
  center of the starburst region, $N_{c} = 3000$ clouds, $v_{adv} = 500$ \kms,
  and $N_{P} = 10^{6}$ particles, although the results are broadly consistent
  with the full parameter space.}
\end{figure}

\begin{deluxetable}{lccc}
\tabletypesize{\scriptsize}
\tablecaption{Model Parameters}
\tablewidth{0pt}
\tablehead{ 
\colhead{Parameters} &
\colhead{Values}
}
\startdata
Geometry of SB & Spherical\\
Radius of SB & 100 pc\\
Molecular Gas Mass & $1 \times 10^{8} \; M_{\odot}$\\
Mean ISM Density $\langle n \rangle$\footnotemark[1] & $486$ cm$^{-3}$\\
Number of Clouds $N_{c}$ & $200 - 3000$\\
Mean Free Path $\lambda_{mfp}$ & $0.5 - 25$ pc\\
Alfv\'{e}n Speed $v_{A}$ & 960 \kms\\
Wind Speed $v_{adv}$ & $0 - 2000$ \kms
\enddata
\footnotetext[1]{Derived from above parameters for an average particle mass of
twice the proton mass.}
\end{deluxetable}

Additionally, a supernova-driven galactic wind is included in the model with a
linear profile and a wind speed of $v_{adv}$ at a height $z = 100$ pc above
the midplane ($v = v_{adv}(z/100$ pc$)$). We vary $v_{adv}$ from 0 to 2000
\kms, with 500 \kms nominally taken as the favored value
\citep{YoastHull2013}. We assume the wind carries away the magnetic field
lines embedded in the hot gas, together with the cosmic rays loaded onto the
field lines, but that the molecular component remains behind.

Since the starburst region of M82 is highly turbulent, we expect the magnetic
field to have a significant random component. Cosmic ray propagation along
tangled magnetic field lines is approximated as a random walk process governed
by a mean free path, $\lambda_{mfp}$, which parameterizes the magnetic
correlation length. As the magnetic geometry is not well known, we vary
$\lambda_{mfp}$ from $0.5$ pc to $25$ pc. A longer value of $\lambda_{mfp}$
(i.e., $\lambda_{mfp} = 50$ pc) was found to yield comparable results to the
$\lambda_{mfp} = 25$ pc case. Due to scattering by short wavelength Alfv\'{e}n
waves generated by the cosmic ray streaming instability, cosmic rays are taken
to travel along the field lines at the Alfv\'{e}n speed, $v_{A}$
\citep{Kulsrud1969}. Given a magnetic field strength $B = 275$ $\mu$G and an
average density of the hot medium $\langle n \rangle = 0.33$ cm$^{-3}$
\citep{YoastHull2013}, the Alfv\'{e}n speed is found to be $v_{A} = 960$ \kms.

We use three methods of injecting the cosmic rays into the starburst
region. First, we consider the simple scenario where all cosmic rays are
injected at the center of the region. Second, in accordance with models of
distributed cosmic ray acceleration such as second-order Fermi acceleration by
interstellar turbulence, we inject the particles randomly throughout the
region. Finally, in agreement with models of point source cosmic ray
acceleration such as first-order Fermi acceleration by supernova shocks, we
inject the particles at randomly chosen supernova shock sites. We choose 30
injection sites due to observational evidence from radio interferometry for
$\sim$ 30 active supernova remnants (SNRs) in M82 \citep[e.g.,][]{Fenech2010}.
One might expect SNRs and star-forming molecular clouds to spatially coincide,
as young, massive stars are not likely to travel far from their place of birth
over the course of their lifetimes. Additionally, recent observations of
$\gamma$-ray emission from molecular clouds associated with SNRs
\citep[e.g.,][]{Aharonian2008, Aharonian2008b, Abdo2010, Abdo2011, Ajello2012}
provide evidence for a spatial correlation between SNRs and molecular
clouds. Thus, for our final injection method, our injection sites are chosen
randomly on spheres of radius $r = 3$ pc centered on clouds. For all cosmic
ray injection methods, we disallow injection inside of clouds due to the
reduced acceleration efficiency expected in cold, dense environments.

The Monte Carlo simulations are run with $N_{P} = 10^{4}$ particles for a
given choice of $N_{c}$, $v_{adv}$, $\lambda_{mfp}$, and cosmic ray injection
mechanism (see the Appendix for a discussion of the number of particles
necessary to achieve convergence). At the beginning of a run, we select the
cosmic rays' initial injection sites as well as a random molecular cloud
distribution, and we map the state of the ISM onto a three-dimensional grid
with a resolution of 0.5 pc. Starting from its injection site, a given
particle executes a random walk governed by a mean free path $\lambda_{mfp}$
broken down into steps of length $l = 0.5$ pc. At each step, the contribution
of advection to the particle motion, the component of the ISM being sampled
(in/out of cloud), and the particle's presence in the starburst region (in/out
of region) are assessed. If a particle is inside of a cloud at the beginning
of a step, the galactic wind does not act on the particle for that step, and
the particle is taken to travel at the speed of light instead of at the
Alfv\'{e}n speed. The particles are taken to free-stream inside of molecular
clouds due to the destruction by ion-neutral damping of Alfv\'{e}n waves
resulting from the streaming instability \citep{Kulsrud1969}. The particle is
scattered when it has traveled a distance equal to its mean free path, and the
process repeats until the particle has traveled out of the region. 

Note that we do not explicitly model energy losses along the particle
trajectories, deferring a discussion of it to Section \ref{sec_disc}. The
cloud column densities ($\sim 5 \times 10^{23}$ cm$^{-2}$, $N_{c} = 3000$
clouds; $\sim 7 \times 10^{24}$ cm$^{-2}$, $N_{c} = 200$ clouds) are well
below the characteristic column densities for energy loss ($N_{E} \sim 2
\times 10^{26}$ cm$^{-2}$, $E = 1$ GeV; $N_{E} \sim 3 \times 10^{25}$
cm$^{-2}$, $E = 1$ TeV) except for very small $N_{c}$ and/or very high proton
energies. This suggests that cosmic rays make several passes though clouds
before undergoing collisional energy losses. See Table 2 for a summary of our
model parameters.

Figure 1 summarizes the sampling behavior of the cosmic rays for a single
representative run of the Monte Carlo code. For each $\lambda_{mfp}$, we show
distributions of the total distance traveled, the total distance traveled in
molecular clouds, and the average density sampled ($\langle \rho_{P} \rangle$)
normalized to the mean density of the ISM in the starburst region ($\langle
\rho_{SB} \rangle$). It is clear from the density distributions that the
particles display a diverse range of sampling behaviors. The extremes of this
behavior are seen in the peaks at $\langle \rho_{P} \rangle / \langle
\rho_{SB} \rangle = 0$, where no clouds are sampled, as well as in the tails
at $\langle \rho_{P} \rangle / \langle \rho_{SB} \rangle \gtrsim 5$, where the
particles sample densities as high as an order of magnitude greater than the
mean density of the ISM. Note that for $\lambda_{mfp} > 1$, a peak appears
near $\langle \rho_{P} \rangle / \langle \rho_{SB} \rangle = 1$. These trends
hint that as we proceed to more precisely quantify the sampling behavior, we
may reasonably expect the particles to roughly sample the mean density of the
ISM over much of parameter space.

\subsection{Emissivity Calculations}\label{sec_emiss}

To quantitatively assess the assumption that cosmic rays sample the mean
density of the ISM in the starburst region and evaluate its effect on
$\gamma$-ray emission, we use our simulation results to construct a function
$\varphi (n/\langle n\rangle) d(n/\langle n\rangle)$, which gives the fraction
of cosmic rays that sample gas density $n$ relative to the mean density
$\langle n\rangle$. The emissivity due to pion production by cosmic rays in
molecular clouds is then given by
\begin{equation}
  \epsilon = \int\varphi\bigg(\frac{n}{\langle n \rangle}\bigg) n_{cr} n\langle
  \sigma_{coll}v \rangle E_{int} \,d\bigg(\frac{n}{\langle n \rangle}\bigg),
\end{equation}
where $n$ and $n_{cr}$ are the number densities of the medium and the cosmic
rays, respectively, $\sigma_{coll}$ is the cross section for pion production,
$v$ is the velocity of the particles, and $E_{int}$ is the energy produced by
the interaction. If the cosmic rays do indeed sample the mean density of the
ISM, the emissivity is given by
\begin{equation}
\epsilon_{0} = n_{cr} \langle n \rangle \langle \sigma_{coll}v \rangle E_{int}.
\end{equation}
Thus, the ratio of emissivities $\alpha$ is given by
\begin{equation}
  \alpha = \frac{\epsilon}{\epsilon_{0}} = \int\varphi\bigg(\frac{n}{\langle n
    \rangle}\bigg) \bigg(\frac{n}{\langle n \rangle}\bigg)
  d\bigg(\frac{n}{\langle n \rangle}\bigg).
\end{equation}
Although $\sigma_{coll}$ is a function of energy, $\alpha$ is independent of
energy for all cosmic rays that satisfy the propagation assumptions given at
the beginning of this section (i.e., the streaming instability results in
propagation at the Alfv\'{e}n speed). $\alpha$ is easily obtained by
appropriately normalizing, binning, and summing over the distributions of
average densities sampled weighted by the densities themselves (see, e.g.,
Figure 1(c)). Thus, we determine $\alpha$ for the range of parameters discussed
in Section 2.1 and seek areas of parameter space where $\alpha$ departs
significantly from unity. See the Appendix for a discussion of our
determination of $\alpha$ for a given choice of $N_{c}$, $v_{adv}$,
$\lambda_{mfp}$, and cosmic ray injection mechanism. The Appendix also details
our estimation of the errors on $\alpha$ that originate from both the finite
number of particles per run as well as the changes in the random cloud
distributions and injection sites from run to run.

\section{Results}\label{sec_results}

\subsection{Emissivity Ratios for all Parameters}\label{sec_allalpha}

\begin{figure}[h]
\epsscale{1.2}\plotone{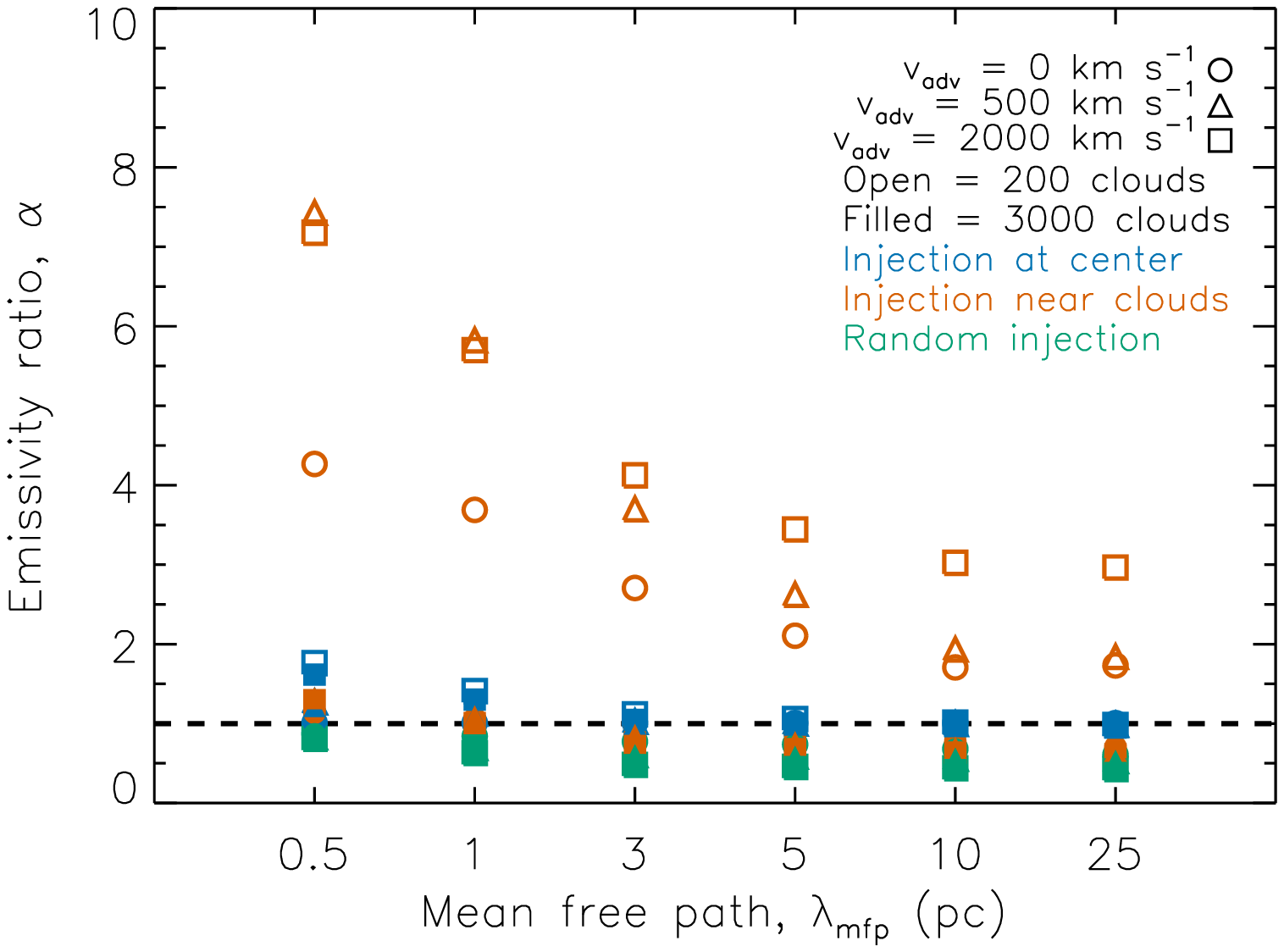}
\epsscale{1.2}\plotone{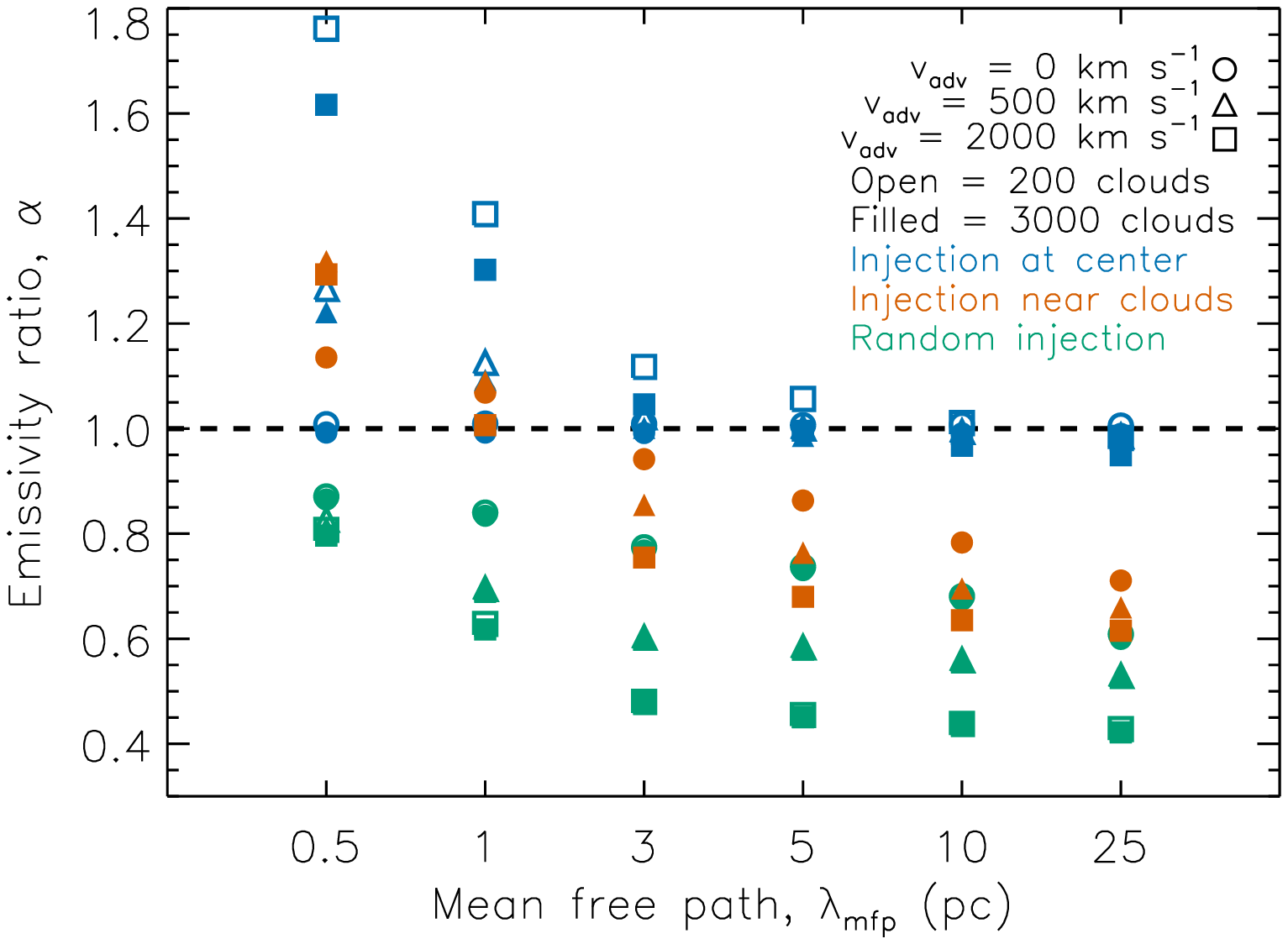}
\caption{Mean emissivities for all combinations of cosmic ray mean free path,
  injection mechanism (indicated by color), three wind speeds (shape), and two
  cloud numbers (fill). The lower panel is a closer view of a portion of the
  upper. It is clear that over the vast majority of parameter space, the
  emissivity ratios are clustered around $\alpha = 1$ and thus the cosmic rays
  roughly sample the mean density of the ISM. However, as discussed in Section
  3.2, cosmic ray injection near the clouds in the limit of small cloud number
  and thus high cloud density results in emissivity ratios that are elevated
  by a factor of a few and the assumption that the cosmic rays sample the mean
  density breaks down. Additionally, cosmic ray injection randomly throughout
  the region results in emissivities that are suppressed by as much as a
  factor of $\sim 2$; however, the suppression of the emissivity at low values
  of $\alpha$ is not as dramatic as its elevation at high $\alpha$.}
\end{figure}

We begin by examining the densities sampled over the entire range of
parameters considered in this study (i.e., all combinations of cosmic ray mean
free path, injection mechanism, two molecular cloud numbers ($N_{c} = 200$,
$3000$ clouds), and three wind speeds ($v_{adv} = 0$, $500$, and $2000$
\kms)). In Figure 2, we see that the cosmic rays roughly sample the mean
density of the ISM over much of parameter space. Overall, we do not achieve
emissivities suppressed below $\alpha = 1$ by a factor of more than $\sim 2$;
we do achieve emissivities elevated above $\alpha = 1$ by a factor of a few;
and these modestly elevated emissivities occur only for cosmic ray injection
near clouds in the limit of small cloud number and thus high cloud
density. This case will be considered separately in Section 3.2.

Although $0.5 \lesssim \alpha \lesssim 1.8$ over much of parameter space,
there are still clear trends relating $\alpha$ to $N_{c}$, $\lambda_{mfp}$,
$v_{adv}$, and cosmic ray injection mechanism. We note that $\alpha$ tends to
increase as $\lambda_{mfp}$ decreases, suggesting that cosmic rays with short
$\lambda_{mfp}$ generally travel along more convoluted trajectories and thus
spend more time sampling the starburst region as well as the clouds that they
encounter. We also note a somewhat greater spread in $\alpha$ at shorter
$\lambda_{mfp}$ that may be due to more frequent scattering allowing for more
varied paths through the starburst region and thus more varied emissivity
outcomes.

For cosmic ray injection near clouds and randomly throughout the region,
$\alpha$ tends to decrease as $v_{adv}$ increases. When injection occurs far
from the midplane, the particle's trajectory is immediately affected by the
wind, and advective losses dominate. However, for injection at the center of
the region and short $\lambda_{mfp}$, $\alpha$ instead increases with
$v_{adv}$. For central injection, the particle initially experiences very weak
wind speeds, and advective losses are negligible. Additionally, for higher
$v_{adv}$, these particles may travel along more efficient trajectories and
thus sample more clouds than for lower $v_{adv}$, where they travel more
convoluted trajectories and are more likely to circumvent clouds.

The cosmic ray injection mechanism also has other effects on the value of
$\alpha$. At a given $\lambda_{mfp}$, emissivities from injection at the
center of the starburst region exceed those of injection randomly throughout
the region, while injection near clouds in the case of $N_{c} = 3000$ clouds
tends to fall between these two limiting cases. It is expected that cosmic
rays injected at the center have the opportunity to sample greater densities
than those injected randomly because they generally spend more time within the
starburst region. Additionally, cosmic rays injected near clouds in the $N_{c}
= 3000$ clouds case are expected to sample greater densities than those
injected randomly because of their initial proximity to clouds. However, as
these cosmic rays also sample lesser densities than those injected at the
center of the region, it appears that this effect is outweighed by the
relatively shorter time spent within the region.

Finally, $N_{c}$ has a clear effect on the value of $\alpha$, albeit one
moderated by injection mechanism. When cosmic rays are injected near clouds
and thus have ample opportunity to encounter dense gas, a smaller number of
denser clouds produces a higher emissivity than a larger number of less dense
clouds. However, when cosmic rays are randomly injected, they have a decreased
likelihood of encountering dense gas; thus, despite the increase in cloud
density, the decrease in cloud number may suppress the emissivity such that
the results are comparable for both a low and high number of clouds.

\subsection{Emissivity Ratios for Injection Near $N_{c} = 200$ Clouds}\label{sec_highalpha}

\begin{figure}[h]
\epsscale{1.2}\plotone{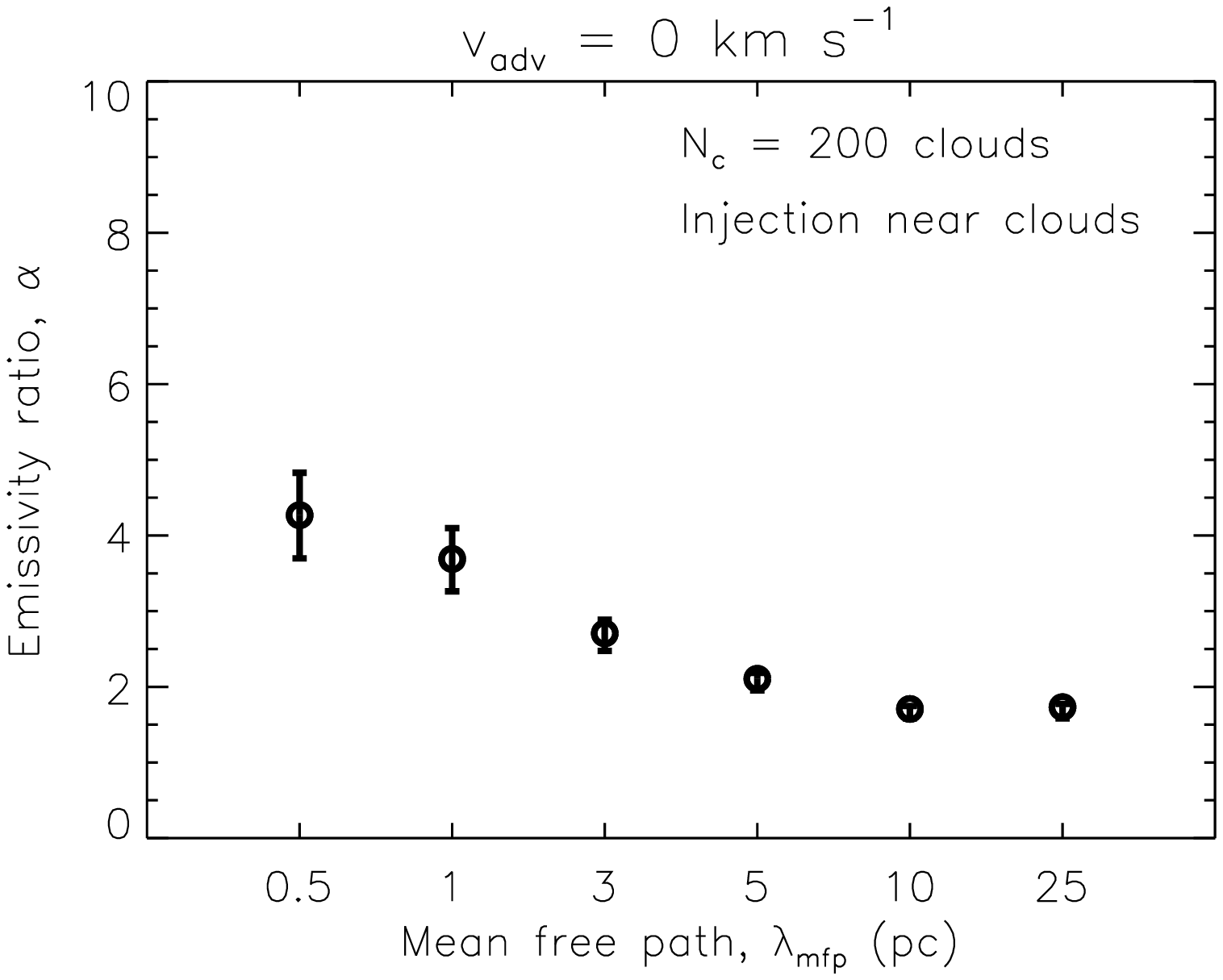}
\epsscale{1.2}\plotone{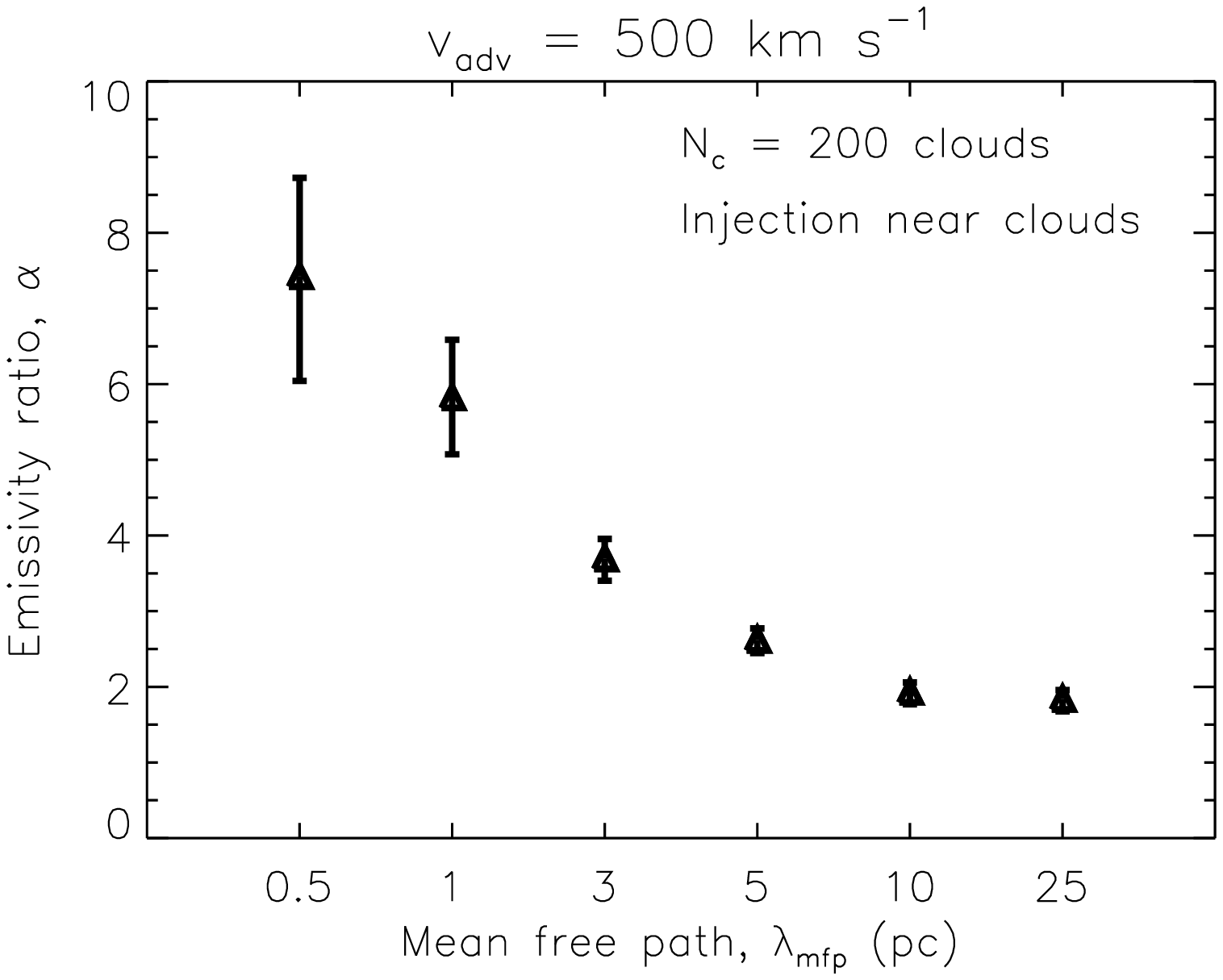}
\epsscale{1.2}\plotone{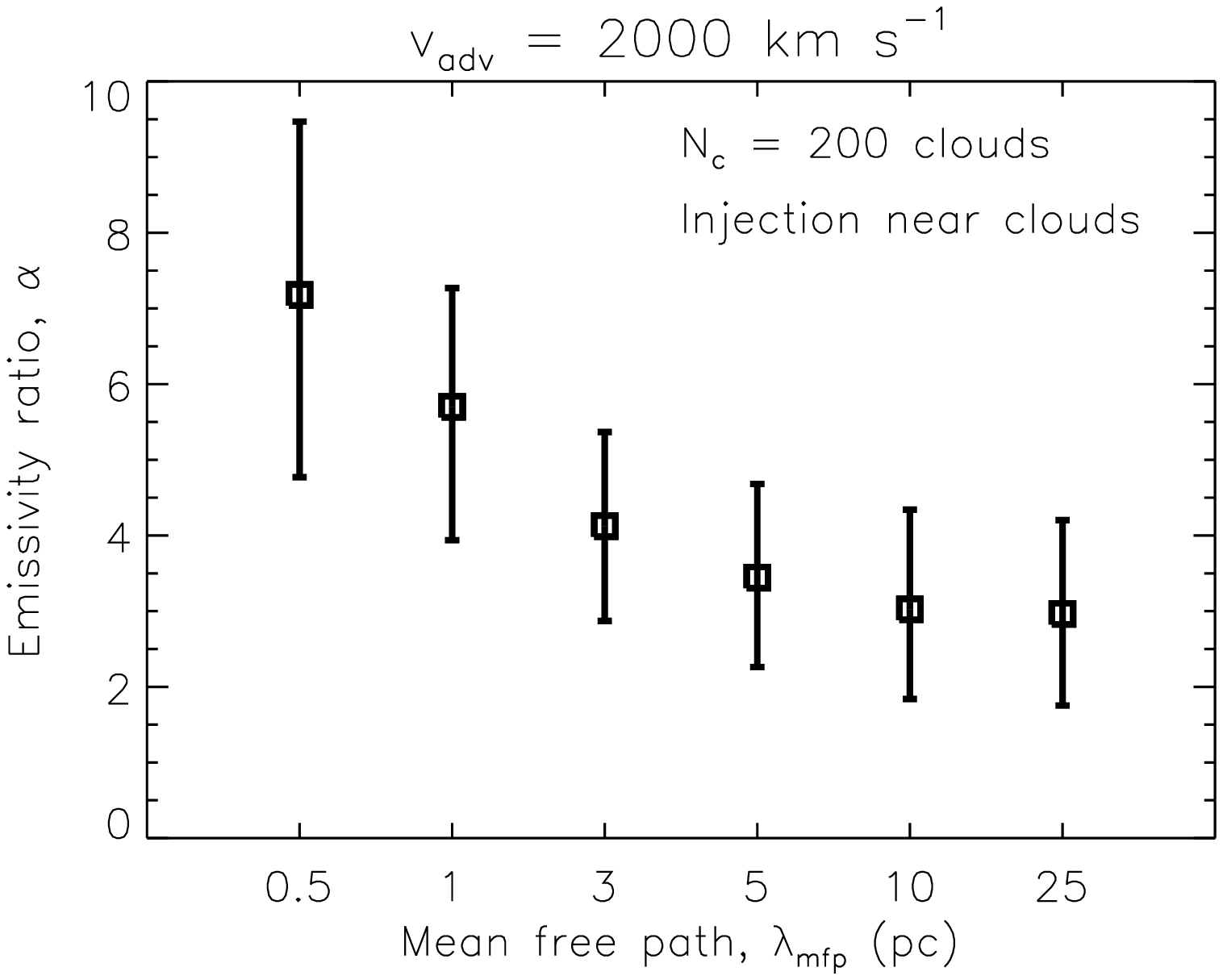}
\caption{Mean emissivities and $68\%$ confidence intervals for cosmic ray
  injection near clouds in the limit of small cloud number ($N_{c} = 200$
  clouds) and high cloud density. In this region of parameter space, we
  achieve emissivities elevated above $\alpha = 1$ by a factor of a few, with
  the highest emissivities achieved with large $v_{adv}$ and small
  $\lambda_{mfp}$.}
\end{figure}

We now discuss the region of parameter space where the cosmic rays sample
the highest densities. As shown in Figure 3, the cosmic rays achieve the
highest $\alpha$ values ($2 \lesssim \alpha \lesssim 7.5$) when injected near
the molecular clouds in the limit of small cloud number ($N_{c} = 200$ clouds)
and thus high cloud density. This result is not surprising, as cosmic ray
injection near high-density molecular gas has the greatest likelihood of
interaction between the particles and the molecular medium. As noted in
Section 3.1, the value of $\alpha$ is highest for short $\lambda_{mfp}$
($\lambda_{mfp} \lesssim 5$ pc), where $\alpha$ varies from $\sim 2$ - $7.5$,
than for longer $\lambda_{mfp}$ , where $\alpha$ is $\sim 2$ - $3$. In
addition to spending more time sampling the starburst region, particles with
short $\lambda_{mfp}$ may also be more likely to enter the clouds accompanying
their injection sites than those that are able to travel more efficiently away.

As is clear in Figure 3, the case of no galactic wind results in $\alpha$
values lower than that of high wind ($v_{adv} = 2000$ \kms) by a factor of
$\sim 2$, while the case of moderate wind ($v_{adv} = 500$ \kms) results in
intermediate emissivities. At short $\lambda_{mfp}$, the cases of high and
moderate $v_{adv}$ become comparable. Particles that are injected between the
galactic plane and their accompanying cloud are likely to be advected into the
cloud in the presence of a wind; the likelihood of this occurring may increase
with $v_{adv}$. Particles with short $\lambda_{mfp}$ also have an increased
likelihood of entering the clouds at their injection sites due to their
inability to efficiently travel away from these sites, rendering the
emissivity less sensitive to $v_{adv}$.

In summary, cosmic ray injection near $N_{c} = 200$ clouds for short
$\lambda_{mfp}$ and moderate to high $v_{adv}$ results in $\alpha$ values that
are modestly elevated above $\alpha = 1$ by a factor of a few. Note that the
uncertainty associated with these emissivities increases dramatically in the
limit of short $\lambda_{mfp}$ and high $v_{adv}$. The former dependence
suggests that particles that experience more frequent scattering travel along
more diverse trajectories and thus have more diverse emissivity
outcomes. Additionally, the latter dependence implies that a higher $v_{adv}$
results in greater populations of particles that are rapidly advected away
(resulting in low emissivities) as well as particles that enter the clouds
near their injection sites (high emissivities). Therefore, although this
region of parameter space results in elevated emissivities, the nature of the
random cloud and injection point distributions is important in determining the
observed emissivity.

We have now seen that cosmic rays injected near $N_{c} = 200$ high-density
clouds sample densities in excess of the mean density of the ISM, while those
injected near $N_{c} = 3000$ lower density clouds do not. Thus, for the case
of injection near clouds, we seek an upper limit on the number of clouds
necessary to achieve elevated emissivities. In Figure 4, we consider $\alpha$
for $v_{adv} = 500$ \kms and a range of $N_{c}$ values for which $200 \leq
N_{c} \leq 3000$ clouds. It is clear that very low values of $N_{c}$ are
required to achieve elevated emissivities. For the shortest $\lambda_{mfp}$,
decreasing $N_{c}$ from $3000$ to $1000$ clouds increases $\alpha$ by only a
factor of $\sim 2$. It is only when $N_{c}$ is decreased to $\sim 400$ clouds
that $\alpha$ becomes elevated by a factor of a few. Thus, in the case of
cosmic ray injection near clouds, a very small cloud number and thus very high
cloud density is required to achieve elevated emissivities.

\begin{figure}[h]
\epsscale{1.2}\plotone{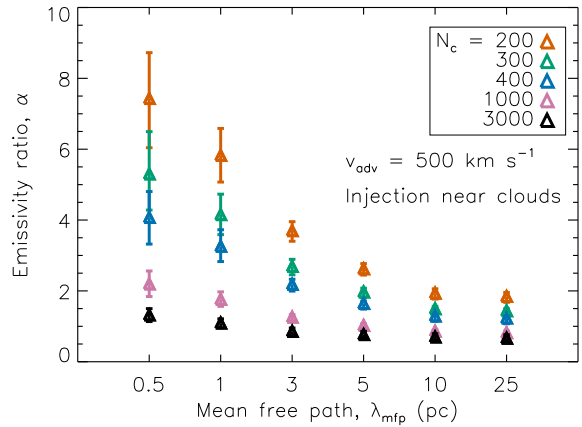}
\caption{Mean emissivities and $68\%$ confidence intervals are strongly
  dependent on the cloud number and thus the cloud density for cosmic ray
  injection near clouds. Though the emissivity is significantly elevated for
  short $\lambda_{mfp}$ in the limit of small $N_{c}$, these emissivities
  decrease very rapidly as $N_{c}$ increases. At $N_{c}$ values greater than
  several hundred, $\alpha$ approaches unity for all $\lambda_{mfp}$.}
\end{figure}

\subsection{Cosmic Rays that Do Not Sample Clouds}\label{sec_nosample}

\begin{figure}[h]
\epsscale{1.2}\plotone{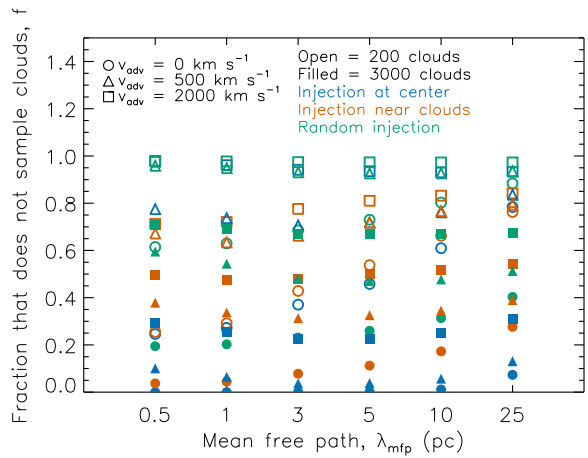}
\caption{Fraction of cosmic rays $f$ that escape the starburst region without
  sampling molecular clouds as a function of $\lambda_{mfp}$. Though the
  cosmic rays approximately sample the mean density of the ISM over much of
  parameter space, they clearly undergo a diverse range of sampling behaviors,
  from near complete sampling of clouds ($f \sim 2 \times 10^{-5}$) to near
  absence of sampling ($f \sim 0.98$). These fractions are determined from
  single $N_{P} = 10^{6}$ particle runs, and will change slightly for
  different distributions of molecular clouds and cosmic ray injection sites.}
\end{figure}

We now consider the fraction of cosmic rays that escape from the starburst
region without sampling molecular clouds at all. This sheds light on the
relationship between the sampling behavior of individual particles and the
sampling of the particle population as a whole. In Figure 5, it is clear that
over the full parameter space considered, the fraction $f$ of cosmic rays that
escape from the starburst region without sampling clouds ranges from near
complete sampling ($f \sim 2 \times 10^{-5}$, $N_{c} = 3000$ clouds, $v_{adv}
= 0$ \kms, injection at the center) to near complete escape ($f \sim 0.98$,
$N_{c} = 200$, $v_{adv} = 2000$ \kms, random injection). As we found that
cosmic rays approximately sample the mean density of the ISM over much of
parameter space, we see that emissivity values of $\alpha \sim 1$ are achieved
over the full range in $f$. Therefore, we may find $\alpha \sim 1$ both when
the vast majority of particles sample roughly the mean density (occurring
mainly for high $N_{c}$ values), as well as when only a small minority of
particles encounter the molecular medium (mainly low $N_{c}$ values).

When we are interested in the mean density sampled by the cosmic ray
population as a whole, such as when evaluating the cosmic ray calorimeter
model, these two broad sampling behaviors can be considered
comparable. However, if large numbers of cosmic rays escape from the starburst
region without losing energy to collisional processes in the molecular medium,
this may have important implications for the galactic environment. It has been
shown that when the cosmic rays are self-confined, meaning that they generate
the waves that trap them, they may contribute to driving a galactic wind
\citep{Breitschwerdt1991, Everett2008}. Thus, we may be more likely to observe
a galactic wind under conditions for which cosmic rays are able to travel
through the hot medium to escape the starburst region without undergoing
energy losses in molecular clouds.

\section{Discussion and Applications}\label{sec_disc}

\subsection{Comparison to a ``Back of the Envelope'' Model}\label{sec_comp}

We now discuss our Monte Carlo simulations in the context of a simple, ``back
of the envelope'' model of cosmic ray sampling of a clumpy ISM. We refer to
the model in Section 4.1 of \citet{YoastHull2013} adapted to account for
diffusive cosmic ray propagation. This model suggests that cosmic rays will
sample the mean density of the ISM if two general conditions are met. First,
the magnetic field lines along which they propagate must pass through a
representative sample of the varied components of the ISM. Second, they must
be able to travel along or diffuse across field lines to encounter these
components before they are advected from the region. To determine whether
these conditions are met, we compare the timescale $\tau_{diff}$ for cosmic
rays to diffuse between clouds to the timescale $\tau_{adv}$ for them to be
advected from the region. We define the former as
\begin{equation}
\tau_{diff} = \frac{l_{c}^{2}}{v_{A}\lambda_{mfp}},
\end{equation}
where $l_{c}$ is the mean distance between clouds and the factor of
$l_{c}/\lambda_{mfp}$ accounts for cosmic ray diffusion with a mean free path
$\lambda_{mfp} < l_{c}$. $l_{c}$ can be approximated as $l_{c} \sim
R/N_{c}^{1/3}$, where $R$ is the radius of the starburst region. We take $l_{c}$
to be the mean minimum distance that cosmic rays must travel to reach a cloud
assuming that there is no spatial correlation between clouds and cosmic ray
injection sites. For this reason, this model is applicable only to cosmic rays
traveling along sufficiently tangled magnetic field lines ($\lambda_{mfp} <
l_{c}$) and to cosmic rays that have been injected either centrally or
randomly in the starburst region.

We define the condition under which cosmic rays will encounter molecular
clouds to be
\begin{equation}
\frac{\tau_{adv}}{\tau_{diff}} \sim 2 \frac{v_{A}}{v_{adv}}
\frac{\lambda_{mfp}}{R} N_{c}^{2/3} > 1,
\end{equation}
where $\tau_{adv} = 2R/v_{adv}$ is the advection timescale for cosmic rays
that experience a mean advecting wind speed of $\sim v_{adv}/2$. Taking $v_{A}
= 960$ \kms, $v_{adv} = 500$ \kms, and $R = 100$ pc, we arrive at a
relationship between $N_{c}$ and $\lambda_{mfp}$ that predicts whether or not
cosmic rays are able to encounter clouds before being advected from the
starburst region:
\begin{equation}
N_{c} > \Bigg(\frac{1}{(0.04 \; \text{pc}^{-1}) \lambda_{mfp}}\Bigg)^{3/2}.
\end{equation}  

We now compare this prediction with the results of our Monte Carlo
simulations. Considering broadly the predictions of Equation (6) for central or
random cosmic ray injection, $v_{adv} = 500$ \kms, and $\lambda_{mfp} <
l_{c}$, this prediction suggests that for all but the shortest
$\lambda_{mfp}$, cosmic rays should encounter dense gas for both $N_{c} = 200$
and $3000$ clouds. For central injection, there is very good agreement between
this prediction and our simulation results, as our $\alpha$ values are indeed
close to unity. For random injection, however, there is only very broad
agreement, as $\alpha$ is instead close to $\sim 0.6$. This discrepancy is not
surprising for the case of random injection, as our simple model does not
account for the effective decrease in the advection timescale for particles
injected considerably closer than $R = 100$ pc to the edge of the region.

Additionally, for $\lambda_{mfp} = 0.5$ pc, Equation (6) suggests that we require
$N_{c} \gtrsim 350$ clouds for the cosmic rays to encounter dense gas. For
central cosmic ray injection, however, we find from our simulations that the
cosmic rays do indeed interact with dense gas for both $N_{c} = 200$ and
$3000$ clouds. For random injection, we again find that the value of $\alpha$
is relatively insensitive to cloud number. This discrepancy is due to our
simple model's inability to account for the effect of increasing cloud density
with decreasing cloud number. In our simulations, we find that despite fewer
cosmic rays sampling clouds in the limit of small cloud number, we still
observe emissivities comparable to those found for larger cloud numbers due to
the increase in cloud density.

Thus, our simple model is useful for broadly considering constraints on the
conditions for cosmic ray sampling of molecular clouds. However, this model
solely suggests whether or not cosmic rays encounter dense gas. We have seen
that the observed emissivity is dependent not only on getting cosmic rays to
the gas, but on additional properties such as the gas density as well. Thus,
for accurately predicting observed emissivities, our full Monte Carlo
simulations are required to account for all relevant subtleties.

\subsection{Implications for Starburst Calorimeter Models}\label{sec_cal}

The starburst calorimeter model suggests that there is a direct relationship
between the energy imparted to cosmic rays by supernovae and the energy lost
by cosmic rays within the starburst region. Thus, there is a relationship
between the supernova rate, and therefore the SFR, and observable emission in
the radio (cosmic ray electrons) and $\gamma$-ray (cosmic ray protons)
regimes. Here, we evaluate the extent to which our simulated starburst region
behaves as a cosmic ray calorimeter under a range of physical conditions by
comparing the particles' confinement timescales $\tau_{C}$ to their energy
loss timescales $\tau_{E}$. Note that as starburst galaxies with properties
like those of M82 are well established to be effective cosmic ray electron
calorimeters \citep[e.g.,][]{YoastHull2013}, we will consider only cosmic ray
protons for the remainder of our analysis.

\begin{figure}[h]
\epsscale{1.2}\plotone{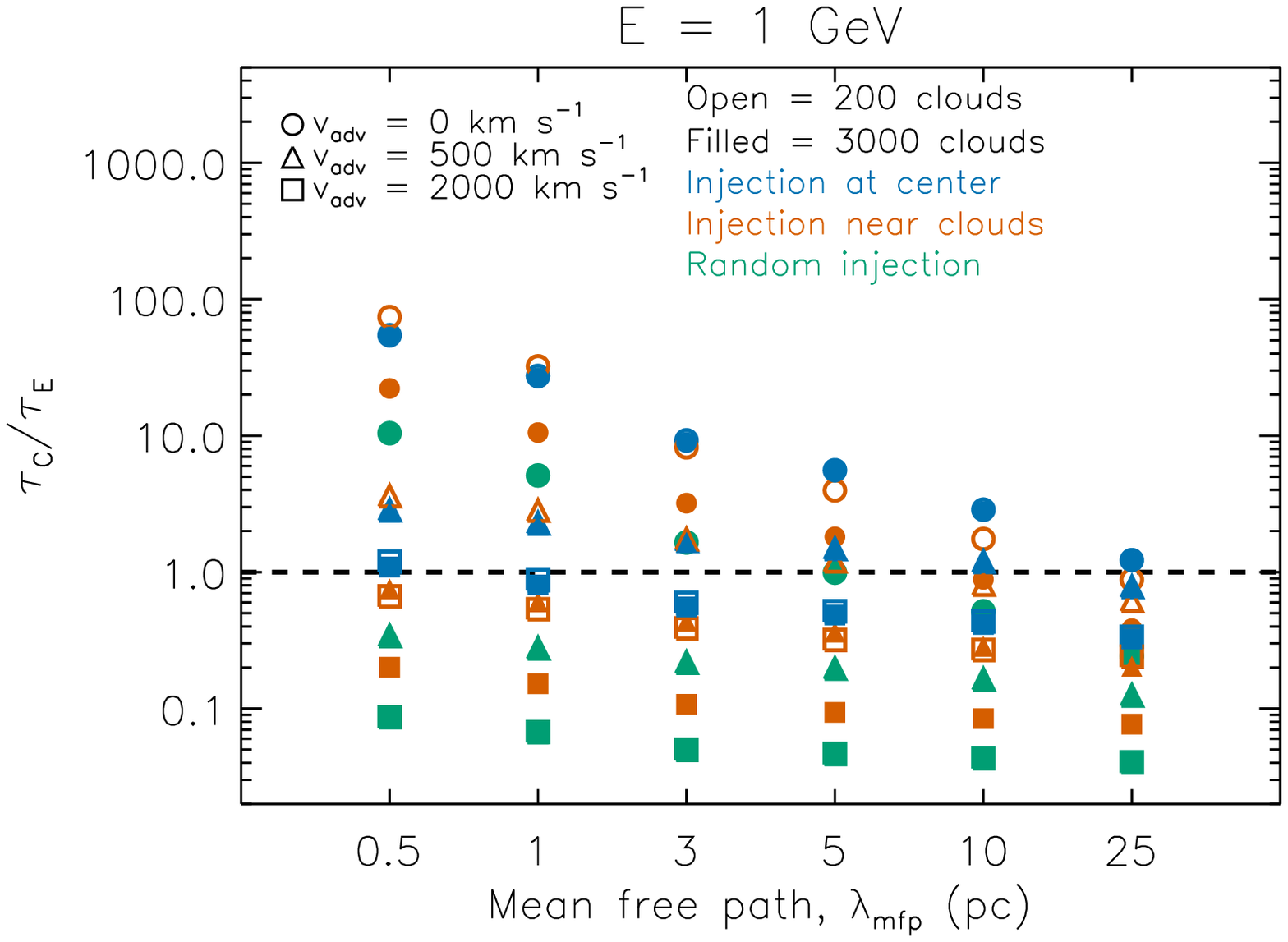}
\epsscale{1.2}\plotone{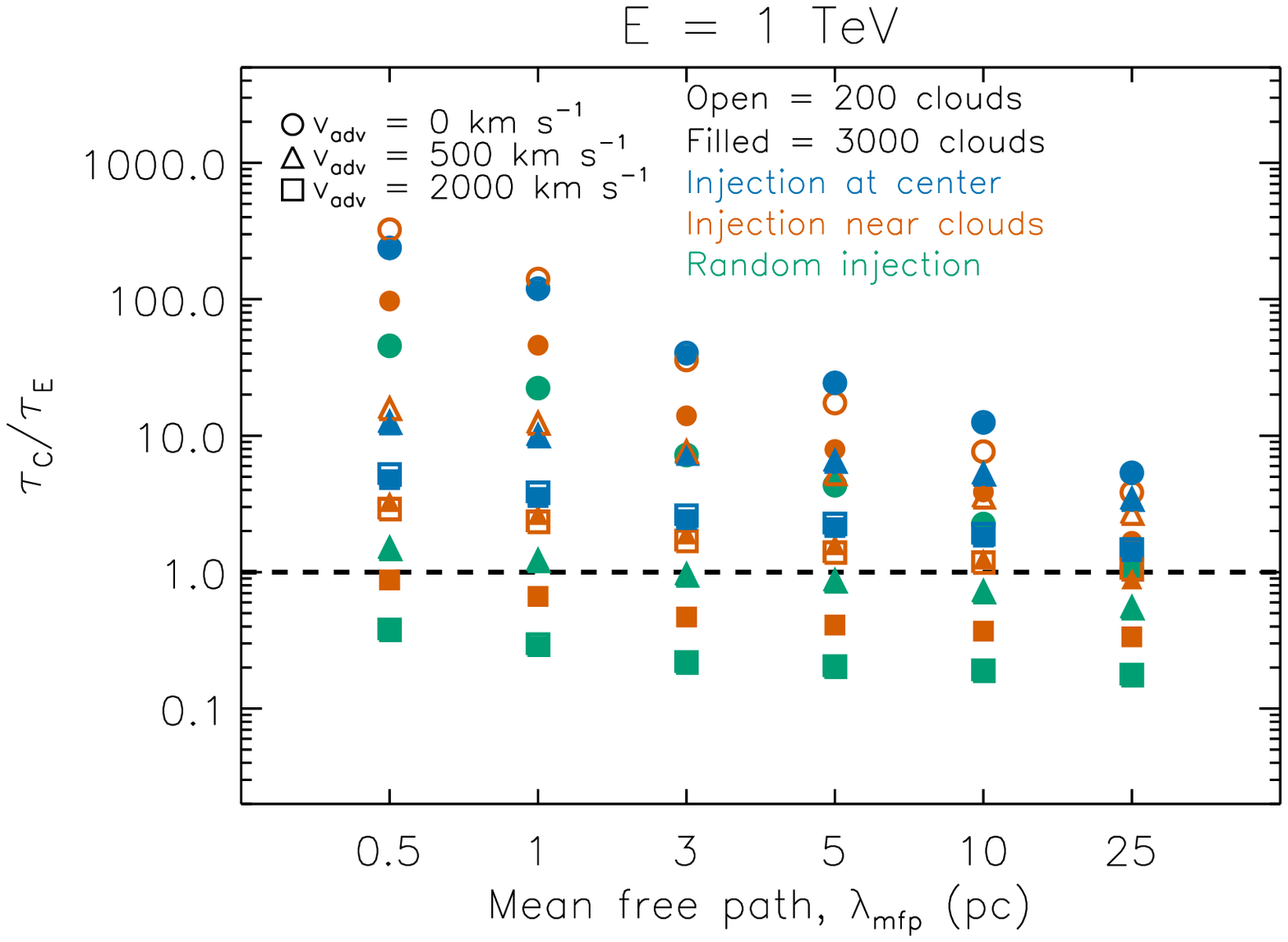}
\caption{To assess whether our simulated starburst region is an effective
  cosmic ray proton calorimeter, we examine the ratio of the confinement
  timescale $\tau_{C}$ to the energy loss timescale $\tau_{E}$. Here, we
  consider $\tau_{C}/\tau_{E}$ for $E = 1$ GeV and $E = 1$ TeV protons for a
  range of conditions on $N_{c}$, $\lambda_{mfp}$, $v_{adv}$, and the cosmic
  ray injection mechanism. Although the starburst region is an effective
  proton calorimeter at low energies ($\tau_{C}/\tau_{E} > 1$, $E \sim 0.1$
  GeV), it may be at best a partial proton calorimeter for the energies
  considered here, where $\tau_{C}/\tau_{E} \lesssim 1$ over non-trivial
  portions of parameter space.}
\end{figure}

The cosmic ray transport equation gives rise to the functional form of
$\tau_{E}$ \citep[e.g.][]{YoastHull2013} and is given by \citep{Longair2011}:
\begin{equation}
  \frac{\partial N(E,t)}{\partial t} = - \frac{\partial}{\partial
    E}\Bigg[\frac{dE}{dt}N(E,t)\Bigg] + Q(E,t) - \frac{N(E,t)}{\tau_{C}}.
\end{equation}
Here, $N(E,t)dE$ is the cosmic ray number density at time $t$ with energies
between $E$ and $E+dE$, $dE/dt$ is the rate at which radiative and collisional
losses decrease a particle's energy, and $Q(E,t)dE$ is the rate at which
particles are injected per unit volume with energies between $E$ and
$E+dE$. Here, $\tau_{C}$ accounts for both diffusive and advective losses and
is assumed to be independent of energy. We may assume that $Q(E)$ and $N(E)$
are of the form $Q(E) = AE^{-\gamma}$ and $N(E) \approx Q(E)\tau(E)$, where
$\tau(E)^{-1} \equiv \tau_{C}^{-1} + \tau_{E}^{-1}$ is the total energy loss
rate. For the steady state, the energy loss rate $\tau_{E}$ is then found to
be
\begin{equation}
  \tau_{E} \equiv -\frac{E}{dE/dt}.
\end{equation}

The energy loss rates for cosmic ray protons are due primarily to ionization
and pion production, the latter of which is dominant above $\sim 1$ GeV
\citep{Schlickeiser2002}. Note that we do not account for the (negligible)
contribution from Coulomb effects. The energy loss rates from both ionization
and pion production are directly proportional to the average density sampled
by the particles, defined as the mean density of the ISM modified by a
multiplicative factor of $\alpha$, $\langle n_{P} \rangle = \alpha \langle
n_{SB} \rangle$. We define the confinement timescale $\tau_{C}$ for a given
$v_{adv}$, $\lambda_{mfp}$, and cosmic ray injection mechanism as the median
time that $N_{P} = 10^{6}$ particles take to escape from the starburst
region. We calculate the ratio of the confinement to energy loss timescales
$\tau_{C}/\tau_{E}$ and seek physical conditions where the calorimeter model
fails to hold ($\tau_{C}/\tau_{E} << 1$).

In Figure 6, we show $\tau_{C}/\tau_{E}$ for $E = 1$ GeV and $E = 1$ TeV
protons over the full parameter space considered. At a given energy, the value
of $\tau_{C}/\tau_{E}$ ranges over more than three orders of magnitude and is
strongly dependent on $v_{adv}$, $\lambda_{mfp}$, and the cosmic ray injection
mechanism. For the lowest energy protons ($E \sim 0.1$ GeV),
$\tau_{C}/\tau_{E} \gtrsim 1$ for all parameters considered, and the starburst
region is an effective proton calorimeter. At moderate proton energies ($E
\sim 1$ GeV), however, $\tau_{C}/\tau_{E}$ only exceeds unity for all
$\lambda_{mfp}$ in the absence of a galactic wind. Additionally,
$\tau_{C}/\tau_{E}$ does not exceed unity for the longest $\lambda_{mfp}$ for
any parameters considered. At higher energies ($E \sim 1$ TeV),
$\tau_{C}/\tau_{E}$ again exceeds unity in all cases except that of large
$N_{c}$, high $v_{adv}$, and injection throughout the region. Thus, for
moderate to high proton energies, our simulated starburst region can only be
deemed a partial proton calorimeter without further knowledge of the physical
conditions in the region. Note that the starburst region would be calorimetric
over larger portions of parameter space if we had made a more generous
estimate of the molecular gas mass contained within the region.

\subsection{The Spectrum of a Cosmic Ray Burst}\label{sec_spec}

We conclude by constructing the spectrum of a cosmic ray burst from a single
supernova explosion and evaluating how the time evolution of the spectrum is
affected by the density sampled by the particles. Although the steady state
solution is generally preferred for the cosmic ray populations of starburst
galaxies, the behavior of a burst of cosmic rays may inform our understanding
of how smaller scale cosmic ray populations evolve over short
timescales. Here, we again consider only cosmic ray protons, as only a small
minority of a supernova's energy is believed to be imparted to cosmic ray
electrons \citep{Blandford1987}.

To construct the spectrum of a burst of cosmic rays, we begin by dropping
both the source and the advection terms from the cosmic ray transport equation;
the advection term will be added later. Thus, we have
\begin{equation}
  \frac{\partial N(E,t)}{\partial t} = \frac{\partial}{\partial E}[b(E)N(E,t)],
\end{equation}
where we define the energy loss rate to be
\begin{equation}
  b(E) = - \frac{dE}{dt}.
\end{equation}
For initial condition $N(E,0) = N_{o}(E)$, one can show that $N(E,t)$ takes
the form:
\begin{equation}
  N(E,t) = N_{o}(E_{o}(E,t))\frac{b(E_{o})}{b(E)},
\end{equation}
where $E_{o}(E,t)$ is the energy at time $t = 0$ of a particle with energy $E$
at time $t$. 

We now restore the advection term to the cosmic ray transport equation:
\begin{equation}
  \frac{\partial N(E,t)}{\partial t} = \frac{\partial}{\partial E}[b(E)N(E,t)]
  - \frac{N(E,t)}{\tau_{C}}.
\end{equation}
We consider the case where $N(E,t)$ assumes the form $N(E,t) =
f(E,t)e^{-t/\tau_{C}}$. By substituting this expression into the transport
equation with advective losses added, we discover that $f(E,t)$ is a solution
of this equation when advective losses are neglected. Thus, we obtain an
expression for $N(E,t)$ accounting for advective losses by multiplying our
previous expression for $N(E,t)$ by a factor of
$e^{-t/\tau_{C}}$. Additionally, we assume that the initial injected spectrum
is a power law $N_{o}(E) = AE^{-\gamma}$ with spectral index $\gamma \gtrsim
2$. Therefore, the spectrum of a burst of cosmic rays has the functional form:
\begin{equation}
N(E,t) = A(E_{o}(E,t))^{-\gamma}e^{-t/\tau_{C}}\frac{b(E_{o})}{b(E)}.
\end{equation}   

To find the initial energy $E_{o}$ given the energy $E$ at time $t$, we solve
for the value of $E_{o}$ that satisfies
\begin{equation}
t = \int_{E}^{E_{o}}\frac{dE^{\prime}}{b(E^{\prime})}.
\end{equation}
This is done by selecting a value for $E_{o}$, numerically integrating
backward, and evaluating the consistency of the resulting $t^{\prime}$ with
the desired time $t$.

Finally, we determine the value of $A$ in order to satisfy:
\begin{equation}
\int_{E_{min}}^{E_{max}}N_{o}(E)E\,dE = \frac{E_{SN}\eta}{V_{SB}},
\end{equation}
where $E_{SN}$ is the energy released by a supernova explosion, $\eta$ is the
fraction of a supernova's energy transferred to cosmic rays, and $V_{SB}$ is the
volume of the starburst region. Therefore, $A$ has the functional form:
\begin{equation}
A = \frac{(\gamma - 2)}{E_{min}^{-\gamma + 2}}\frac{E_{SN}\eta}{V_{SB}},
\end{equation}
where we take $\gamma = 2.1$, $E_{min} = 0.1$ GeV, $E_{SN} = 10^{51}$ erg,
$\eta = 0.1$, and $V_{SB} \sim 10^{62}$ cm$^{3}$. Note that we take the
cosmic ray acceleration timescale in a supernova shock to be $\tau_{accel}
\lesssim 10^{4}$ yr $<< \tau_{C}$, and thus $\tau_{accel}$ can be taken to
be effectively instantaneous.

\begin{figure}[h]
\epsscale{1.2}\plotone{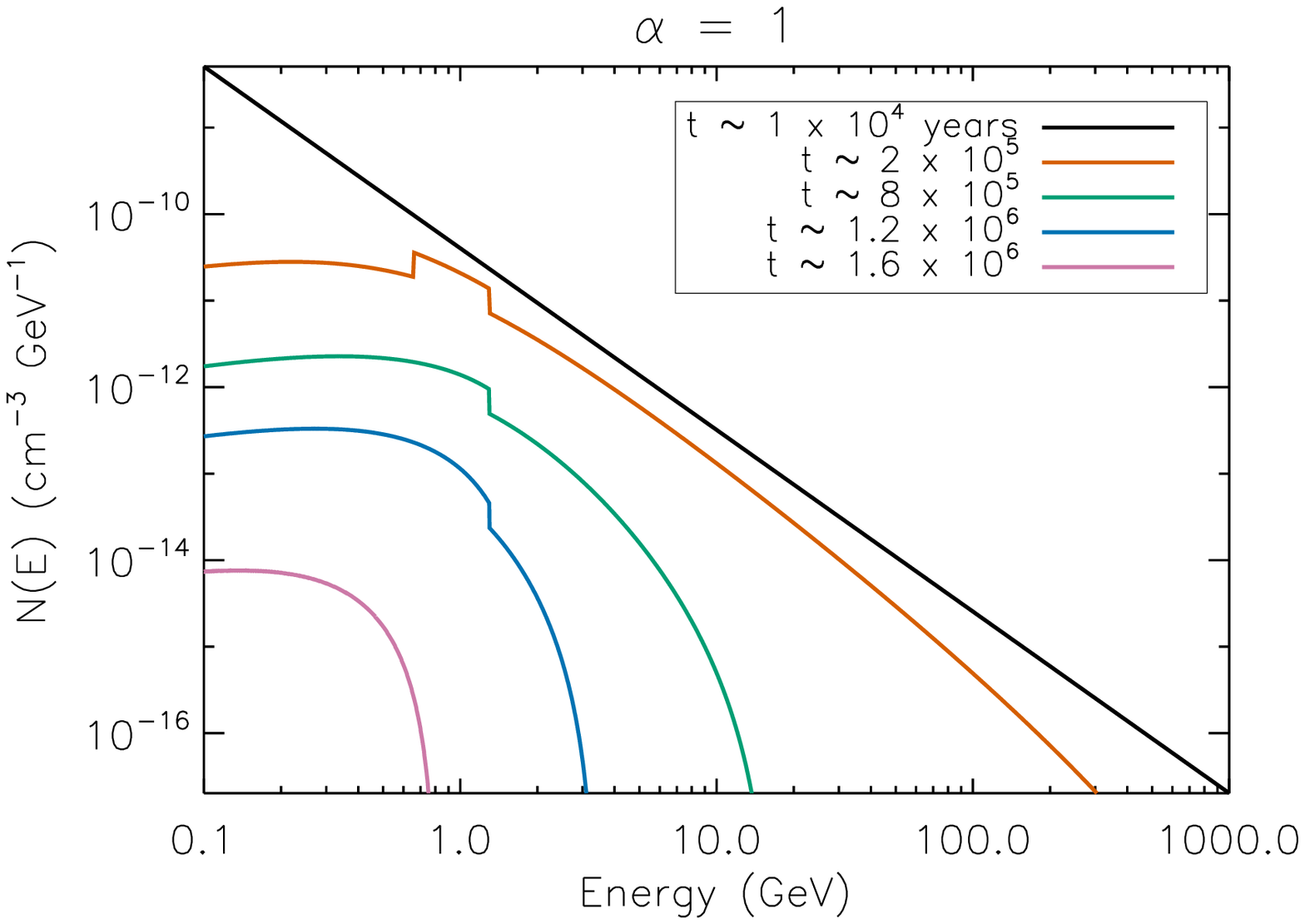}
\epsscale{1.2}\plotone{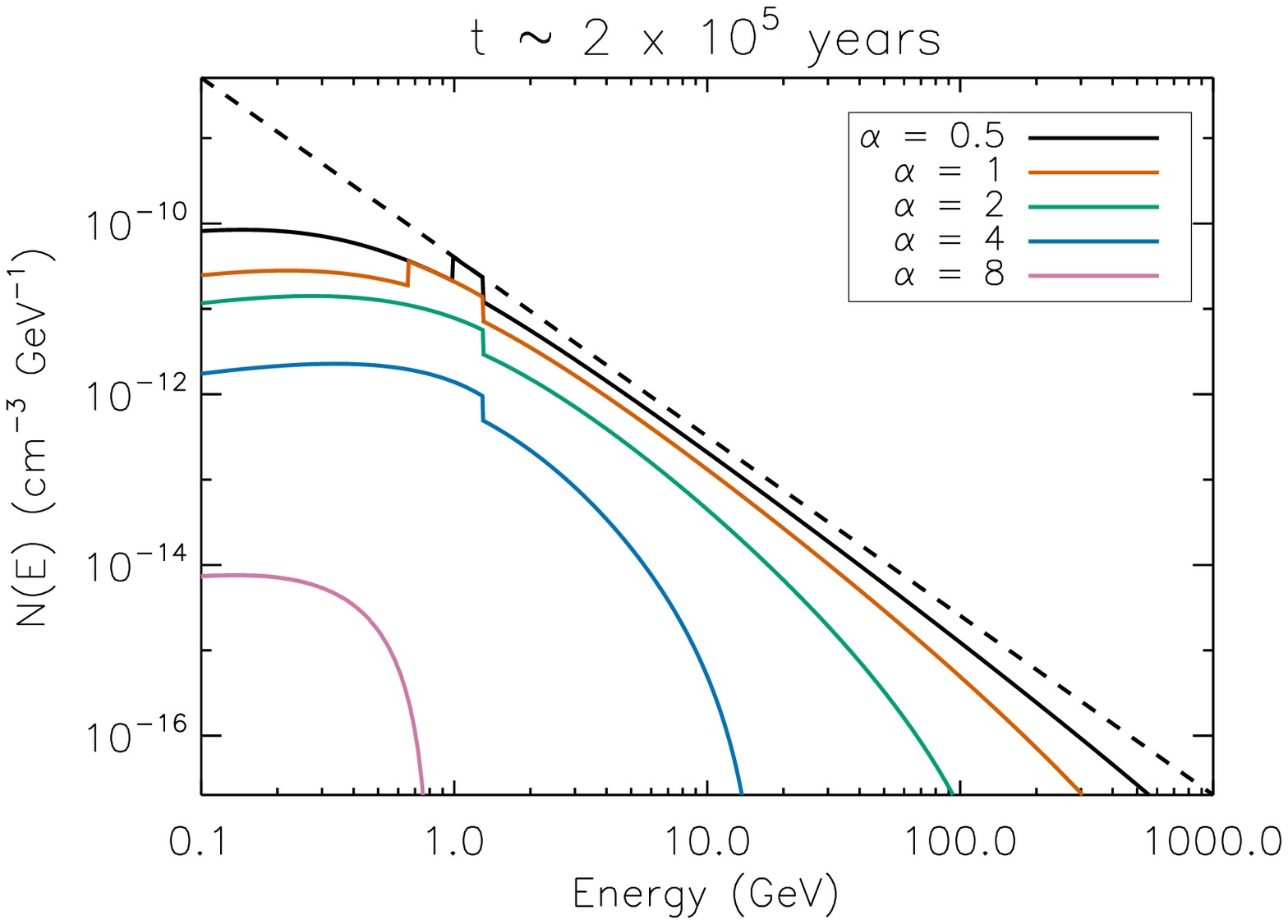}
\caption{In the upper panel, we show the time evolution of the spectrum of a
  burst of cosmic ray protons from a single supernova explosion under the
  physical conditions of M82. Here, we take $\alpha = 1$ and $\tau_{C} =
  10^{6}$ yr, values consistent with much of the parameter space
  considered. Other confinement times can be easily accounted for by
  appropriately adjusting the factor of $e^{-t/\tau_{C}}$ in Equation (13). The
  rapid evolution of the spectrum with time suggests that cosmic ray bursts
  are short lived in these environments. In the lower panel, we show the
  spectrum of a cosmic ray burst at time $t \sim \tau_{accel}$ (dashed line)
  as well as at $t \sim 2 \times 10^{5}$ yr (solid lines) for a range of
  values of $\alpha$. It is clear that cosmic ray bursts that sample densities
  greater than the mean density of the ISM by a factor of a few have
  dramatically shortened lifetimes.}
\end{figure}

The time evolution of the spectrum is shown in the upper panel of Figure 7 for
$\alpha = 1$ and $\tau_{C} = 10^{6}$ yr. This choice of $\alpha$ and
$\tau_{C}$ is consistent with much of the parameter space considered. As in
Section 4.2, the energy loss rates for cosmic ray protons are calculated
according to \citet{Schlickeiser2002}. It is evident that the spectrum evolves
rapidly with time; on timescales comparable to the confinement time, the
number density decreases by several orders of magnitude at all energies. This
decline is most dramatic at high energies ($E \gtrsim 10$ GeV), where the
spectrum steepens sharply by $t \sim 10^{6}$ yr due to extreme energy loss
rates as well as the initial lack of high-energy particles.

In the lower panel of Figure 7, we show the spectrum at time $t \sim 2 \times
10^{5}$ yr for a range of densities sampled. It is clear that increasing
the value of $\alpha$ by a factor of a few rapidly decreases the lifetime of
the particle population. For example, by comparing the top and bottom panels
of Figure 7, we see that the spectrum of a cosmic ray burst with $\alpha = 1$
that has evolved for $t \sim 1.6 \times 10^{6}$ yr is identical to the
spectrum of a burst with $\alpha = 8$ that has evolved for only $t \sim 2
\times 10^{5}$ yr.

Overall, the spectrum of a burst of cosmic rays evolves rapidly under the
physical conditions of a starburst galaxy like M82, and thus the cosmic ray
population produced by a single supernova explosion under these conditions is
a short-lived phenomenon. However, a conservative estimate of the supernova
rate in M82 is $\sim 0.07$ yr$^{-1}$ \citep{Fenech2008}, and thus we can
expect that a supernova explosion will occur in M82 every $\sim 15$ yr. The
evolved cosmic ray spectra shown in Figure 7 will therefore never be made
manifest, and instead will be regularly replenished by newly injected
particles. However, in other star-forming environments where the supernova
rate may be very low, such as extreme dwarf galaxies or OB associations, these
spectra may indeed be manifested due to a lack of replenishing particles.

\section{Conclusions}\label{sec_conc}

In the interest of understanding how cosmic rays sample the clumpy ISM in a
starburst environment, we have undertaken Monte Carlo simulations of cosmic
ray sampling of molecular clouds under the physical conditions of the
archetypal starburst galaxy M82. Here, we briefly review the results of our
study:

\begin{itemize}
\item Cosmic rays sample roughly the mean density of the ISM ($\alpha \sim 1$)
  for a wide range of assumptions about the number of molecular clouds, the
  galactic wind speed, the extent to which the magnetic field is tangled, and
  the cosmic ray injection mechanism. A value of $\alpha \sim 1$ is consistent
  with models of the observed $\gamma$-ray spectrum of M82
  \citep{YoastHull2013}.
\item Cosmic rays sample densities a factor of a few higher than the mean
  density ($2 \lesssim \alpha \lesssim 7.5$) in the case of a small number of
  dense molecular clouds, injection close to the clouds, and a highly tangled
  magnetic field.
\item The fraction of cosmic rays that escape from the starburst region
  without sampling molecular clouds ranges from $10^{-5} \lesssim f \lesssim
  0.98$ for physical conditions that yield $\alpha \sim 1$. This suggests that
  although the sampling behavior of the cosmic ray population as a whole is
  largely independent of the physical conditions, the behavior of individual
  particles is not.
\item Our simulated starburst region is at least a partial cosmic ray proton
  calorimeter, and appears to be a complete calorimeter at all proton energies
  for no galactic wind and short cosmic ray mean free path.
\item We construct the time-dependent spectrum of a burst of cosmic rays, and
  demonstrate that the spectrum evolves rapidly under the physical conditions
  of M82. However, these spectra are only made manifest in environments with
  very low supernova rates ($< 10^{-5}$ yr$^{-1}$) where the spectra are not
  continuously replenished with energetic particles.
\end{itemize}
Though we have illustrated several applications here, cosmic ray sampling of a
clumpy ISM may be applied to a wide range of science goals seeking to
understand the relationship between the state of the multi-phase ISM, the star
formation processes, and the cosmic ray populations of starburst galaxies.

\acknowledgments We gratefully acknowledge the support of NSF AST-0907837 and
NSF PHY-0821899 (to the Center for Magnetic Self-Organization in Laboratory
and Astrophysical Plasmas). We thank Benjamin Brown, Sebastian Heinz, Dan
McCammon, and Joshua Wiener for helpful discussions, as well as Masataka Okabe
and Kei Ito for supplying the colorblind-friendly color palette used in this
paper (see \url{fly.iam.u-tokyo.ac.jp/color/index.html}). This work has made
use of NASA's Astrophysics Data System.

\appendix

\emph{Simulation Convergence.} We use a modified bootstrapping technique
\citep{Efron1979} to determine the minimum number of particles necessary to
achieve convergence in our Monte Carlo simulations. For each cosmic ray mean
free path, injection mechanism, and parameter extrema (i.e., $N_{c} = 200$ and
$3000$ clouds, $v_{adv} = 0$, $500$, and $2000$ \kms), we resampled an $N_{P}
= 10^{6}$ particle run with replacement $10^{4}$ times for a range of sample
sizes ($10^{2} \leq N_{P} \leq 10^{4}$). We calculated a new value of $\alpha$
for each selected sample and fit the resulting distributions in $\alpha$
with Gaussian profiles to determine their standard deviations. As we are
primarily interested in areas of parameter space where $\alpha$ departs
significantly from unity, our purposes are easily served by uncertainties in
$\alpha$ on the order of $\sim 10\%$. In Figure 8, we show an example of
the resulting distributions in $\alpha$ with a range of sample sizes.

\begin{figure}[h]
\epsscale{1.0}\plotone{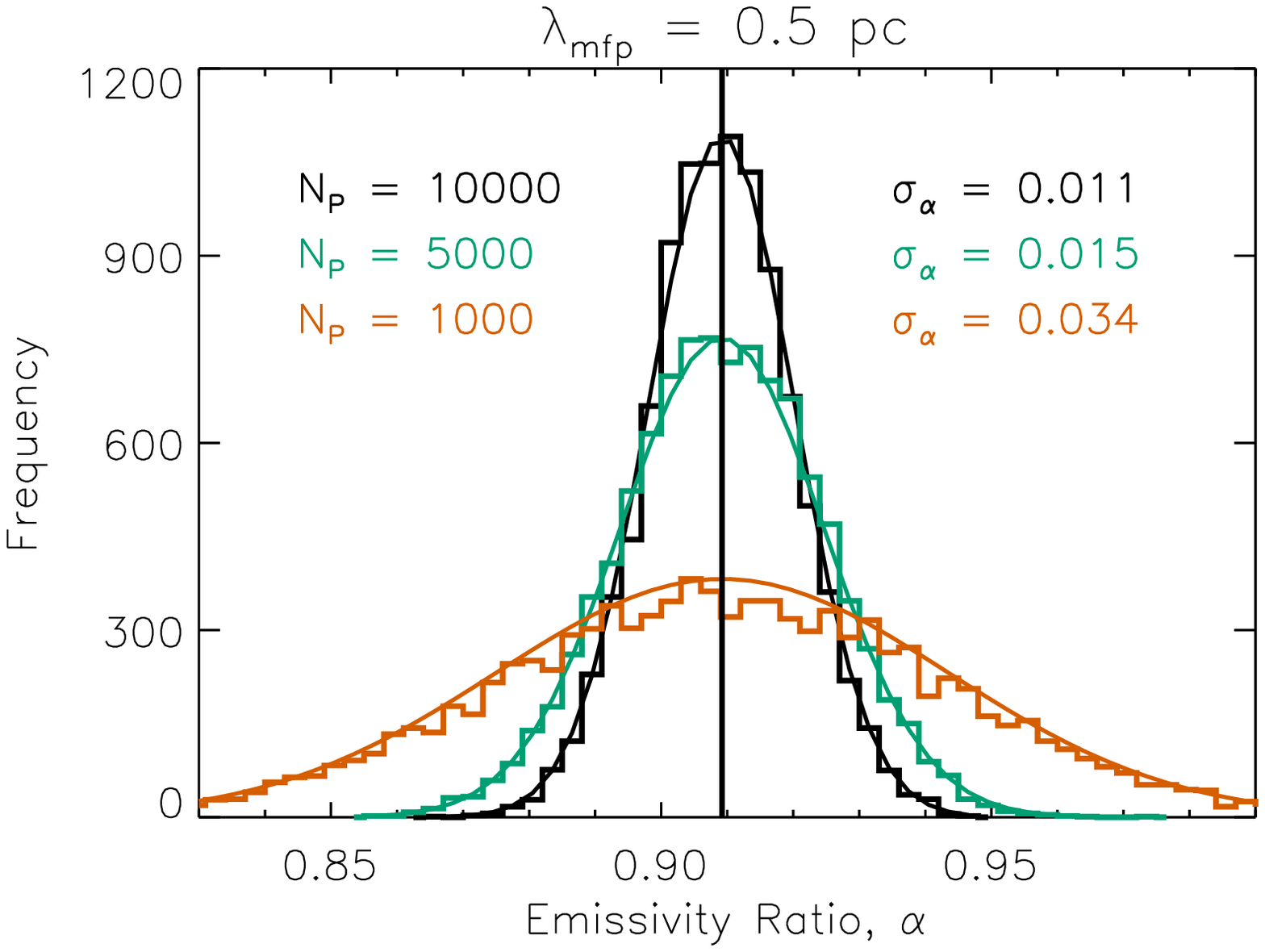}
\epsscale{1.0}\plotone{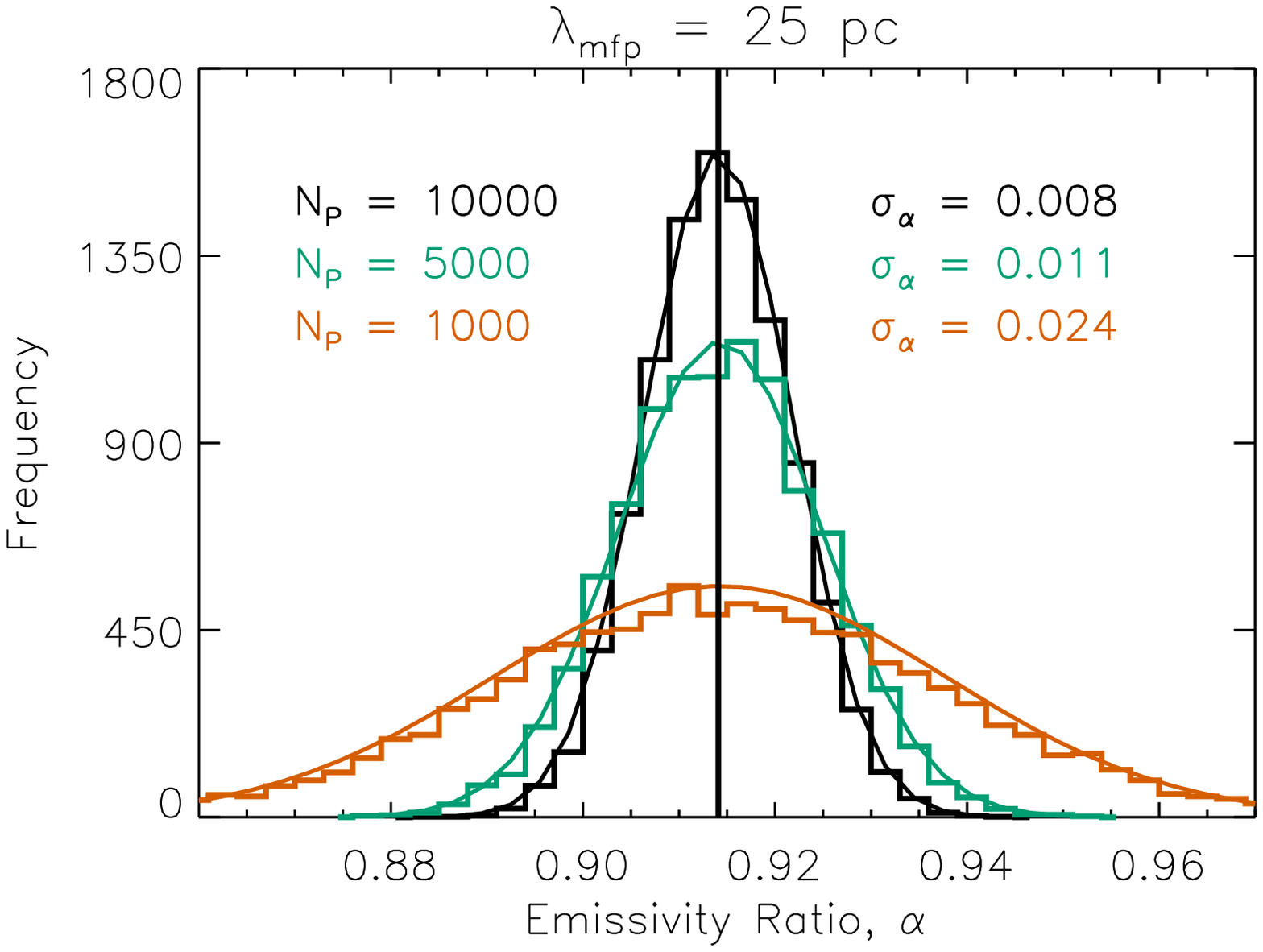}
\caption{To determine the number of particles necessary to achieve convergence
  in $\alpha$, we resampled runs with $N_{P} = 10^{6}$ particles with
  replacement $10^{4}$ times with a variety of sample sizes ($10^{2} \leq
  N_{P} \leq 10^{4}$ particles). Here, we show example resulting distributions
  in $\alpha$ for sample sizes of $N_{P} = 10^{3}, 5 \times 10^{3}$, and
  $10^{4}$ particles. The solid curves are Gaussian fits to these
  distributions, and the solid line indicates the value of $\alpha$ obtained
  from all $N_{P} = 10^{6}$ particles. Across the full parameter space, $N_{P}
  = 10^{4}$ particles are sufficient to achieve $\alpha$ values with
  accompanying uncertainties that are small ($\sigma_{\alpha} \lesssim 13\%$)
  compared to the large changes in $\alpha$ ($\Delta \alpha \gtrsim 1$) that
  we seek. This simulation was run with $N_{c} = 3000$ clouds, $v_{adv} = 500$
  \kms, and injection at the center of the starburst region, though the
  results are generally representative of the full parameter space ($0.3\%
  \leq \sigma_{\alpha} \leq 13\%$).}
\end{figure}

For a sample size of $N_{P} = 10^{4}$ particles, the spreads of the resulting
Gaussian distributions ($\sigma_{\alpha}$) range from $\sim 0.33\%$ (for
$N_{c} = 3000$ clouds, $v_{adv} = 0$ \kms, and central injection) to $\sim
13\%$ (for $N_{c} = 200$ clouds, $v_{adv} = 2000$ \kms, and random
injection). Decreasing the sample size to $N_{P} = 5000$ particles increases
$\sigma_{\alpha}$ by $\sim 40\%$, and to $N_{P} = 1000$ particles increases
$\sigma_{\alpha}$ by more than $200\%$.  Additionally, the case of $N_{c} =
200$ clouds and injection randomly throughout the region results in skewed
distributions with significant tails toward high emissivities for $N_{P} =
1000$ and $5000$ particles, although the skew is negligible for $N_{P} =
10^{4}$ particles. Thus, we select $N_{P} = 10^{4}$ particles as the number of
particles necessary to achieve convergence, because this sample size allows
$\sigma_{\alpha}$ to remain below or comparable to $\sim 10\%$ and our
resampled distributions to remain Gaussian across the full parameter space.

\emph{Error Estimation.} The generally dominant source of uncertainty in
$\alpha$ for a given choice of $N_{c}$, $v_{adv}$, $\lambda_{mfp}$, and cosmic
ray injection mechanism is associated with the changes in the molecular cloud
and injection site distributions from run to run. To quantify this
uncertainty, we ran repeated runs varying only the cloud distribution (for
central injection) or the cloud and injection site distributions (for the
other injection methods) until sufficiently sampled emissivity distributions
were obtained (i.e., $\sim 250$ runs). As shown in Figure 9, the shape and
spread of the emissivity distributions are dependent on the cosmic ray
injection mechanism. Random injection results in fairly Gaussian distributions
with modest spreads ($\sigma_{\alpha} \sim 0.01$ - $0.1$), while the other
injection methods result in slightly to severely skewed distributions with
sparsely populated tails at high emissivities. These latter injection
mechanisms also result in significant spreads ($\sigma_{\alpha} \sim 0.03$ -
$1.0$, injection at center; $\sigma_{\alpha} \sim 0.05$ - $2.5$, injection
near clouds). The mean emissivities and $68\%$ confidence intervals obtained
from these distributions are the emissivities and error bars reported in this
work.

\begin{figure}[h]
\epsscale{1.0}\plotone{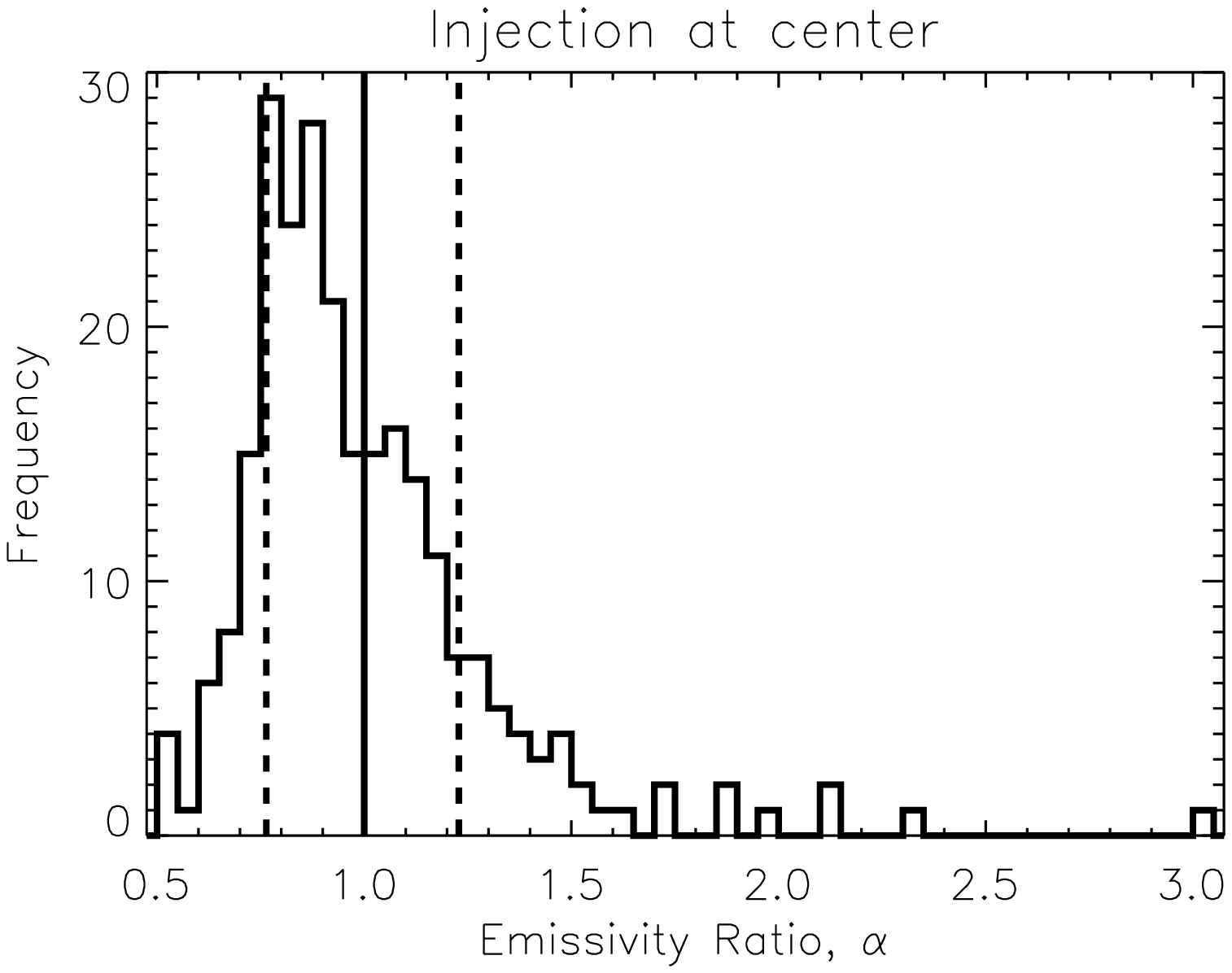}
\epsscale{1.0}\plotone{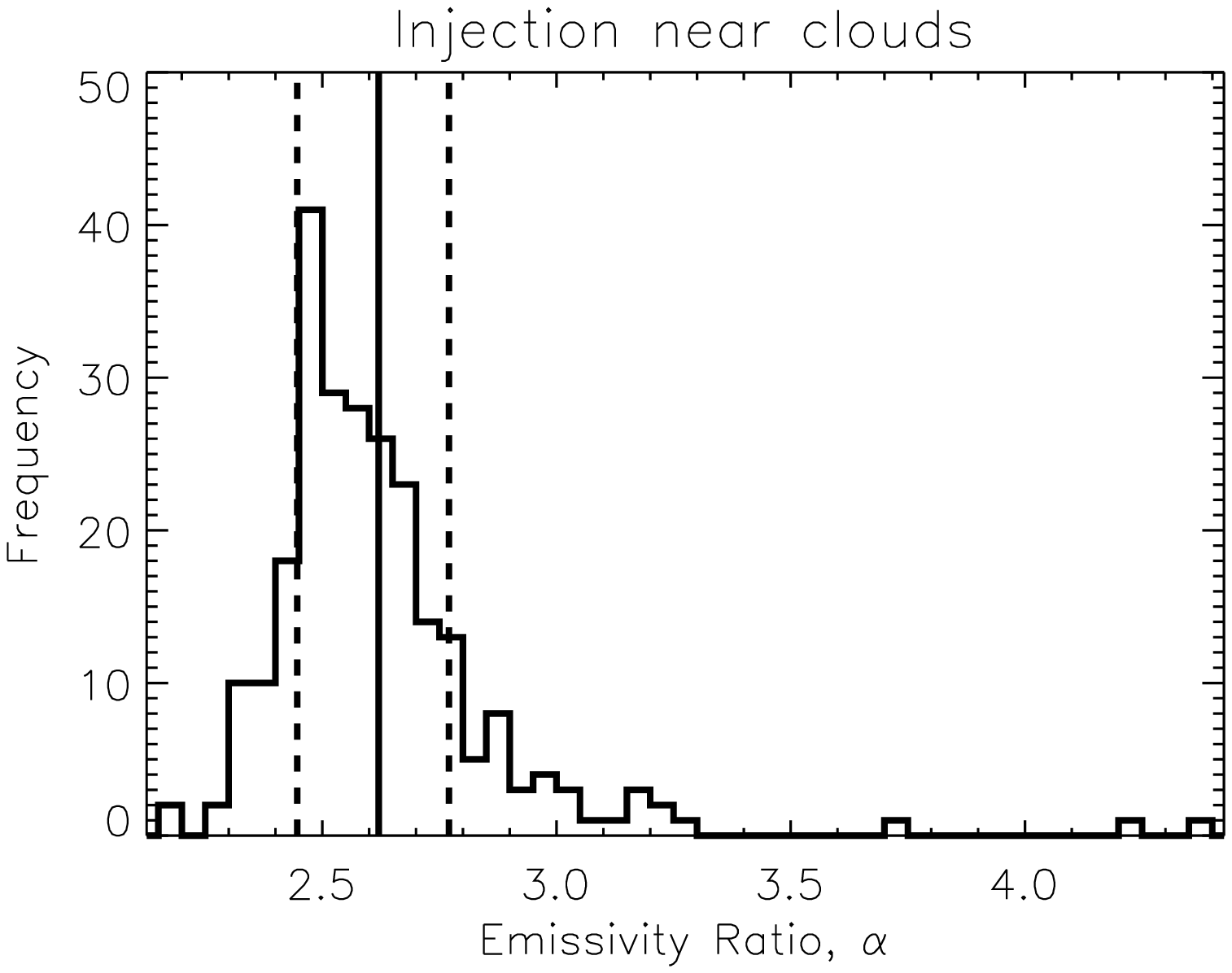}
\epsscale{1.0}\plotone{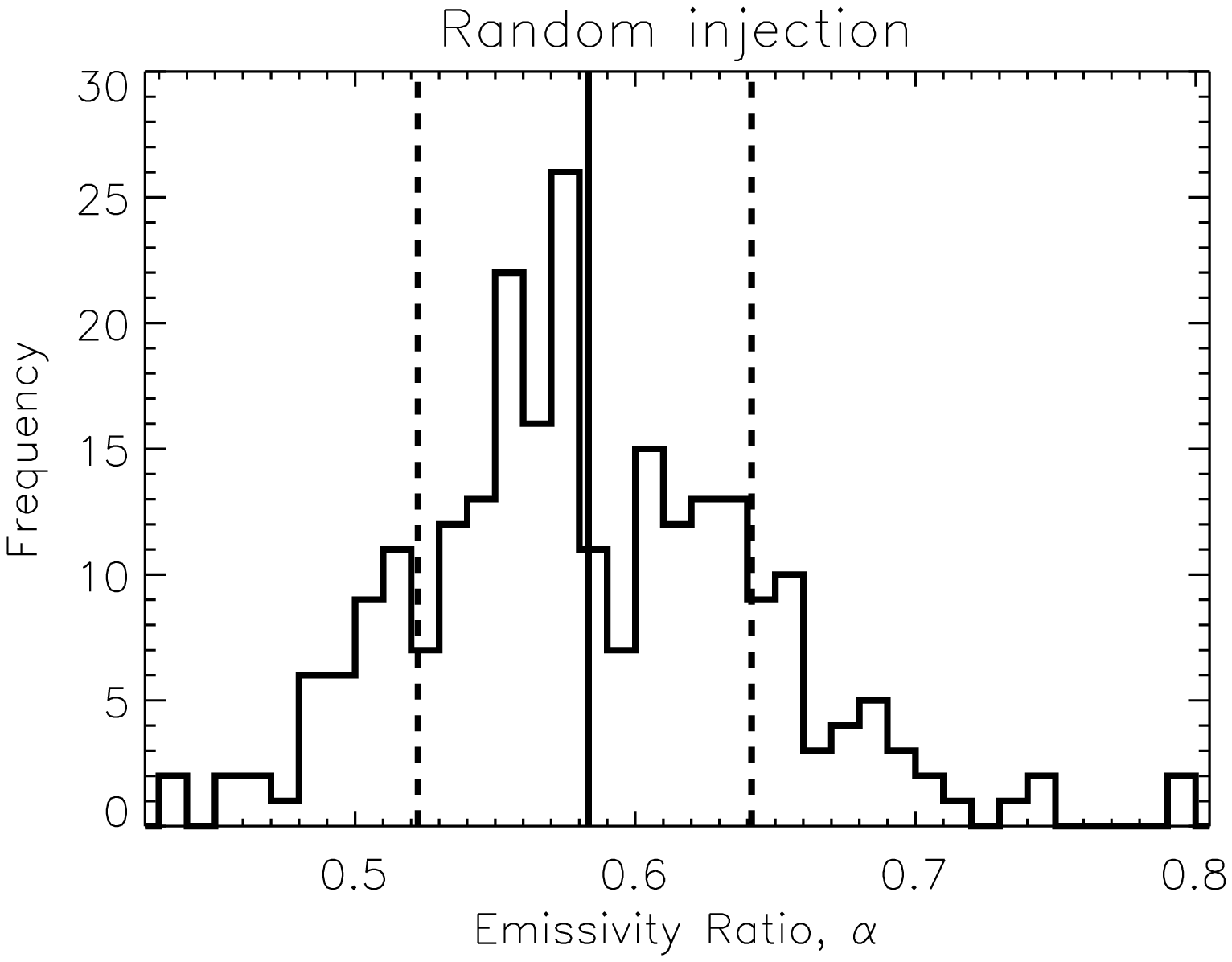}
\caption{Sample emissivity distributions for each injection mechanism used to
  obtain the value of $\alpha$ and the corresponding uncertainty associated
  with the changes in the cloud and injection site distributions from run to
  run. The solid line indicates the mean of the distribution, and the dashed
  lines the $68\%$ confidence intervals. While random cosmic ray injection
  results in fairly Gaussian distributions, it is clear that injection at the
  center and near clouds result in distributions that are non-Gaussian with
  sparsely populated tails at high emissivities. These simulations have $N_{c}
  = 200$ clouds, $\lambda_{mfp} = 5$ pc, and $v_{adv} = 500$ \kms. Note that
  the apparent asymmetry in the area enclosed by the $68\%$ confidence
  intervals in the top and middle panels is due to the choice of binning.}
\end{figure}

\bibliographystyle{apj}

\end{document}